\documentclass{aa}
\usepackage[T1]{fontenc}
\usepackage{graphicx}
\usepackage{txfonts}
\usepackage[T1]{fontenc}
\usepackage{subcaption}
\usepackage{placeins}
\usepackage[table]{xcolor}
\usepackage{algorithm}
\usepackage{algpseudocode}
\usepackage{amsmath}
\usepackage{siunitx}
\usepackage{multirow}
\usepackage{makecell}
\usepackage{booktabs}
\usepackage{natbib}
\usepackage{comment}

\DeclareRobustCommand{\VAN}[3]{#2}
\let\VANthebibliography\thebibliography
\def\thebibliography{\DeclareRobustCommand{\VAN}[3]{##3}\VANthebibliography}

\title{Multi-branch classification of diffuse cluster radio emission
       from the LOFAR two-metre sky survey}

\author{Markus Bredberg\inst{1}\corrauth{markus.bredberg@epfl.ch}
   \and Emma Tolley\inst{1}\email{emma.tolley@epfl.ch}}

\institute{Institute of Physics, Laboratory of Astrophysics,
   École Polytechnique Fédérale de Lausanne (EPFL),
   Observatoire de Sauverny, Versoix, 1290, Switzerland}

\date{Received 12/05/2026; accepted 24/07/2026}

\begin{document}

\abstract
  {Galaxy clusters sometimes host synchrotron radiation on scales of $\sim$\SI{100}{kpc} to $\sim$\SI{1}{Mpc}, with a surface brightness only a few times the image noise. This diffuse cluster radio emission is a sensitive probe of magnetic fields and intracluster medium dynamics, but disentangling the underlying physical processes requires statistically large samples spanning a wide range of cluster masses, dynamical states, and redshifts, together with a sufficient sensitivity to low-surface-brightness emission.}
  {We explore two techniques for improving the detection of diffuse emission in galaxy cluster images, relative to a baseline classifier: the scattering transform (ST) and squeeze-excitation (SE) attention.}
  {We integrated an ST encoder into a dual-branch classifier (DualSSN) and a scattering network (ScatterNet). We incorporated SE attention into the DualSSN and dual-branch convolutional neural network (DualCSN). These classifiers were then benchmarked against a simple convolutional neural network (CNN), across ten image pre-processing configurations and three cropping strategies. Performance was evaluated on small labelled datasets from the second data release of the LOFAR Two-metre Sky Survey overlapping with the Second Planck catalogue of Sunyaev-Zel'dovich sources (LoTSS-DR2/PSZ2).}
  {Alongside the multi-branch approach with SE and ST, cropping the image to a fixed number of telescope beams and \textit{uv} tapering (smoothing to a coarser angular resolution) improve classification performance, while stacking multiple pre-processed versions of an image does not.}
  {Scattering-transform-based multi-branch architectures with beam-normalised cropping are a promising direction for diffuse emission classification in the SKA era.}


\titlerunning{Multi-branch classification of diffuse cluster radio emission}
\authorrunning{Bredberg \& Tolley}
\maketitle
\nolinenumbers

\section{Introduction}

Galaxy clusters are the largest gravitationally bound structures in our Universe, and contain  up to a few thousand galaxies. These galaxies move and interact within the ionised intracluster medium (ICM), which has temperatures of $10^{7-8}\,$K, a particle density of $\sim10^{-3}\,$cm$^{-3}$, and contains magnetic fields of approximately $\mu$G strength \citep[e.g.][]{Clarke01, Bruggen12, Johnston15, Stuardi21}. When galaxy clusters interact, stochastic turbulent re-acceleration, diffusive shock acceleration, or hadronic collisions produce highly energetic cosmic rays. These acceleration mechanisms produce spectra following a power-law distribution at frequencies where the synchrotron source is optically thin \citep{pacholczyk70}. Electrons can reach GeV energies and produce synchrotron radiation at frequencies from $\sim$\SI{10}{MHz} to $\sim$\SI{1}{GHz}, where higher-frequency emission comes from higher-energy particles.
Since the radiative loss of a particle increases with its velocity, lower frequencies display more powerful emission originating from less energetic particles and older events. Typical spectral indices of around $\alpha \approx -1 $ to -1.5 are thus characteristic of aged, non-thermal synchrotron emission, and its precise value constrains the energy spectrum and cooling time of the underlying electron population \citep{Feretti12, vanWeeren19}.

The diffuse cluster radio emission takes different forms depending on the underlying physical processes. Radio halos arise from synchrotron radiation produced by the turbulent re-acceleration of energetic particles from cluster-cluster interactions and trace the centrally located ICM. Radio relics are the afterglow of the ICM shock waves of cluster mergers. They are more elongated, polarised, and located in the periphery of clusters. For reviews of diffuse radio emission in galaxy clusters, readers can refer to \cite{vanWeeren19} and \cite{wittor23}. 

Since synchrotron radiation probes ultrarelativistic particles in magnetic fields and cluster-cluster interactions energise these particles, the existence and morphology of diffuse cluster radio emission contain information about the cluster merger history and early magnetic fields. However, the morphology, spectrum, polarisation, and spectral index are determined by a combination of poorly constrained processes, including particle acceleration mechanisms, magnetic-field amplification, projection effects, and temporal evolution during mergers \citep{Govoni04, vanWeeren19,  Giovannini20, Osinga21, Tevlin25}. Disentangling these contributions requires statistically large samples spanning a wide range of cluster masses, dynamical states, and redshifts, as well as a sufficient sensitivity to low-surface-brightness emission. For current surveys of cluster-scale diffuse sources, the LOFAR Two-metre Sky Survey \citep[LoTSS;][]{shimwell17} has detected definite or candidate diffuse emission in 99 out of 309 (\SI{32}{\%}) clusters \citep{shimwell22, botteon22}. In the MeerKAT Galaxy Cluster Legacy Survey (MGCLS), 55 out of 103 (\SI{53}{\%}) show diffuse emission \citep{knowles22, Kolokythas25}. A study of non-Planck galaxy clusters in the second data release of LoTSS (LoTSS-DR2), probing lower cluster masses, identified tens of additional diffuse sources \citep{Hoan22}.

The number of diffuse cluster samples is expected to improve as radio astronomy enters the upcoming data-intensive era. The LOw-Frequency ARray \citep[LOFAR;][]{Haarlem13} is generating \SI{7.6}{PB} of data from the 3451-hour-long LoTSS, which is being stored in the currently \SI{22}{PB} large J\"ulich LOFAR Long-term Archive \citep{manzano24}. The upcoming Square Kilometre Array (SKA) observatories SKA-Low \citep[in Australia;][]{Labate22} and SKA-Mid \citep[in South Africa;][]{Swart22} are predicted to archive \SI{700}{PB} per year.\footnote{https://www.skao.int/en/explore/big-data} 
This means that many new sources of diffuse cluster radio emission will be discovered, at a lower surface brightness and at a higher redshift.
The SKA1 All-Sky continuum survey is expected to discover $\sim$4000 diffuse emission sources at z$<1$ \citep{Wagg21}. Since exabyte-scale storage is expensive, SKA visibility data will only be provided in exceptional circumstances \citep{ska2025}, and visibility-based deblending of smaller sources and faint extended cluster emission will not be possible. 

The automated detection of diffuse cluster emission will be essential in the SKA era.
Machine learning can be used to extract features of diffuse emission and label existing images \citep{norris16, ndungu23}. \cite{Gheller18} developed a convolutional neural network \citep[CNN;][]{AlexNet12}, named COSMODEEP, capable of detecting patches containing synthetic diffuse cluster radio emission with $\sim$\SI{90}{\%} accuracy. Building on this task, \cite{Stuardi24} trained a U-Net, named RadioUNet, on cosmological simulations and fine-tuned it on LoTSS to successfully recover diffuse emission morphology, and they identified clusters with diffuse emission with \SI{73}{\%} accuracy. This model led to a detection of super-cluster-scale diffuse radio emission, which was previously missed by humans \citep{Stuardi25}. \cite{Sanvitale25} used the same method, but with the convolutional encoder replaced with a vision transform encoder, creating a TransUNet named TUNA. This model identified diffuse radio emission at the LOFAR high-band antenna sensitivity limit with a resolution four to six times coarser than input images. With the help of images generated by a denoising diffusion probabilistic model, \cite{Mishra24} trained a CNN to detect radio halos from the GaLactic and Extragalactic All-sky Murchison widefield array (GLEAM) survey with $95.93\pm1.57\,\%$ accuracy. The trained classifier also correctly identified nine out of the twelve known halos from MGCLS and five out of the eight known halos from the Planck Sunyaev–Zel'dovich Catalogue 2 (PSZ2) \citep{planck15, bahk24}.

Most high-performance classifiers are supervised and therefore rely on labelled data. This is a problem when datasets contain only a few hundred images per class. For instance, in LoTSS-DR2, only 207 clusters are labelled, 178 of which contain a known redshift \citep{botteon22}. One way to adapt machine learning methods to small datasets is to introduce certain interpretable analytical tools, such as the scattering transform \citep[ST;][]{Mallat12, Bruna12} and the squeeze-excitation (SE) attention \citep{Hu17}. 

This paper evaluates multi-branch CNNs, with the ST and SE attention, on small datasets of radio cluster images from LoTSS-DR2.
Specifically, four classifying models are trained and tested on galaxy cluster images with and without diffuse emission. Each classifier combines convolutions, the scattering transform, and SE attention to different degrees. A simple CNN serves as the baseline, and ScatterNet, which replaces learned convolutions with a fixed scattering transform, provides a non-trainable feature-extraction baseline. Two dual-branch architectures, DualCSN and DualSSN, then let us isolate the individual contributions of the dual-branch structure, the scattering transform, and SE attention to classification performance, as detailed in Section~\ref{sec:models}.

The paper is organised as follows. Section~\ref{sec:scattering_transform} and Section~\ref{sec:squeeze_excitation} explain the ST and SE. In Section~\ref{sec:models}, the classifiers are detailed. The dataset is introduced in Section~\ref{sec:data_and_processing}, and the training procedure in Section~\ref{sec:training}. Section~\ref{sec:results} presents the results which are then discussed in Section~\ref{sec:discussion}. Conclusions are presented in Section~\ref{sec:conclusions}.

\section{Methods}

\subsection{Scattering transform} \label{sec:scattering_transform}

The ST is a hierarchical signal processing technique that captures invariant and stable representations of data \citep{Mallat12}. The method iteratively applies Morlet wavelet transforms followed by modulus and averaging operations, resulting in a multi-scale, multi-orientation representation. Each wavelet is specified by scale parameter $j$ measuring pixel scale $2^j$, and an orientation parameter $l$ measuring a direction $\frac{l}{L}\pi$. These parameters range from $0$ to $J-1$ and $L-1$, respectively, creating $\binom{J}{m} L^m$ scattering coefficients per spatial location, for a given order $m$. Mathematically, given an input signal $I$, the ST is computed as:
\begin{equation}
S_m(j_1, l_1,..., j_m, l_m) = ||| I \star \Psi_{j_1, l_1} | \star \ldots | \star \Psi_{j_m, l_m} | \star \Phi \label{eq:scattrans}
\end{equation}
where $\Psi_{j, l}$ denotes the wavelet transform operator at scale $j$ and orientation $l$, $\star$ represents the convolution operation, $\Phi$ is a low-pass filter necessary for the wavelet transform to satisfy the Littlewood-Paley condition and thereby be stable and invertible, and $|\cdot|$ denotes the modulus operation.

Conceptually, the ST can be interpreted as a non-trainable CNN because it iteratively convolves the input image with Morlet wavelets. The first order of scattering coefficients are scale-invariant features that capture edge-like structures and texture orientations, akin to descriptors such as scale-invariant feature transform \citep[SIFT;][]{Lowe99, Bruna12}. The higher-order scattering coefficients, on the other hand, provide deeper invariants that encode more complex patterns within the signal. These higher-order terms effectively capture interactions between different scales and orientations, enabling the representation to discriminate between image fields that have similar Fourier power spectra but differ in higher-order correlations. For non-sparse images, most energy is captured in the first two orders, allowing for good performance of relatively simple scattering networks \citep{Bruna12, Anden14}. 

The ST has been used in astrophysics to capture non-Gaussian features of cosmological fields, often outperforming contemporary state-of-the-art CNNs for similar tasks \citep{Cheng20, Cheng24}. \cite{Saydjari21} reduced the number of scattering coefficients to less than 100 from which they inferred physical properties of the interstellar medium and synthesised realistic fields. Non-Gaussian statistical characteristics have also been extracted with the ST to denoise Planck polarisation observations \citep{Blancard21}, to probe HI emission in the cold neutral medium \citep{Lei23}, for cosmological parameter estimation \citep{Allys19, Allys20, Cheng20, Cheng21, Cheng24}, and in more general astrophysical fields with the first four order moments in a scattering spectrum \citep{Morel22, Cheng23, Mousset24}. For classification tasks, the ST has produced state-of-the-art results for Modified National Institute of Standards and Technology (MNIST) and the United States Postal Service (USPS) handwritten digit datasets \citep{Bruna12}. \cite{Tolley24} demonstrated stable performance on radio galaxies, which are more complex than MNIST digits, with few parameters ($\sim$ 11,000) and noisy data. 

\subsection{Squeeze-excitation attention} \label{sec:squeeze_excitation}

A second useful component in small CNNs is the SE block \citep{Hu17}. It re-weights the channels of the convolutional feature maps which helps the network focus on relevant features. An SE block is a small side-branch linear network that operates on the channel-wise spatial average of the features of that convolutional layer. The squeeze operation extracts channel statistics through global average pooling. The excitation operation produces channel weights for adaptive recalibration via a two-layer feed-forward network with a rectified linear unit (ReLU) \citep{Hinton10} after the first layer and a sigmoid activation after the second. This self-attention mechanism has been shown to improve classification performance on existing CNNs \citep{Hu17}, proven central for attention gates \citep{Oktay18} and convolutional block attention modules \citep{Woo18}, and seen many applications in astrophysics, including gravitational waves identification \citep{Li25} and extraction of cosmological parameters \citep{Farieta24}. 

\subsection{Neural network architectures} \label{sec:models}

\begin{figure}
    \centering
    \includegraphics[width=\hsize]{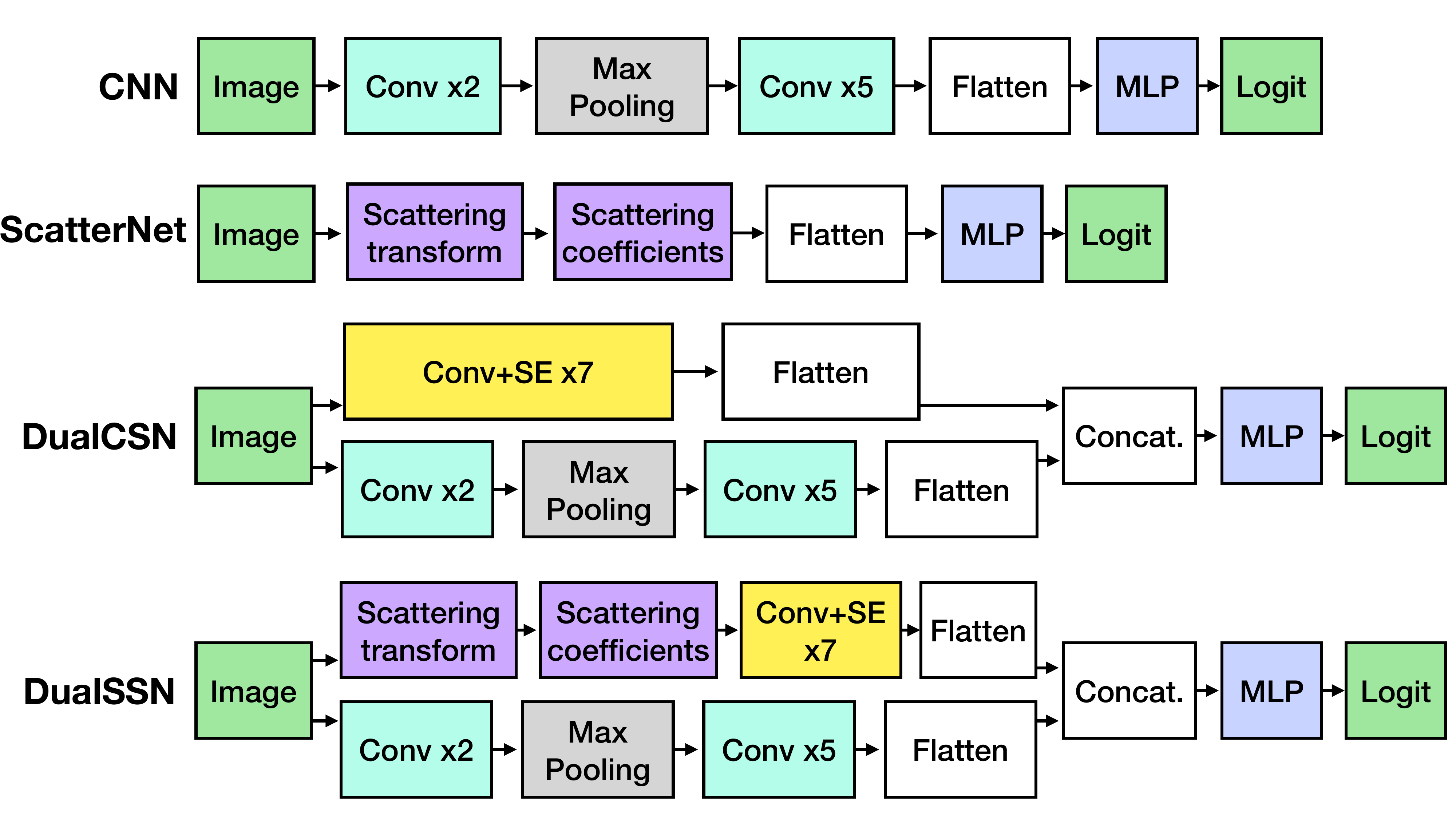}
    \caption{Schematic view of the four different classifier models used in this work.}
    \label{fig:architectures}
\end{figure}

We compared four classifier architectures of increasing structural complexity, as shown in Figure~\ref{fig:architectures}. The first architecture is a CNN and is described in Table~\ref{tab:CNN}. It consists of six convolutional blocks: two blocks with a convolutional size of $3\times3$, one block with $5\times5$, and four stride-2 blocks with $3\times3$. The convolutional blocks are followed by two linear layers. In total, the architecture consists of 69,234 parameters. More parameters would likely overfit on the small dataset \citep{hastie09, goodfellow16}. In addition, the network uses progressive dropout as regularisation \citep{Srivastava14}.

The second model, \emph{ScatterNet}, is the ST connected to a simple multi-layer perceptron (MLP), following \cite{Tolley24}. Its structure is analogous to that of the CNN with the ST instead of convolutions. The ST was calculated with a maximum scale $J=2$ and number of orientations $L=12$ to the second order. This configuration produces a large amount of scattering coefficients and has performed well in previous classification tasks \citep{Kinakh21}. A larger value for $L$ would result in more parameters than the current 1,976,534, as shown in Table~\ref{tab:ScatterNet}. To manage the large number of scattering coefficients, we explored two alternative MLP replacements: a convolutional reduction block analogous to the final four layers of the CNN, and a single-branch version of the DualSSN. Neither produced comparable results, so the simple MLP was retained.

The third and fourth architectures are both dual-branch CNNs. The motivation for a dual-branch design was that different feature extractors tend to make different errors: one branch may be sensitive to large-scale morphology, while the other captures fine-grained texture. Combining two feature extraction modes can thus suppress individual failure modes in a similar manner to ensemble averaging \citep{Lakshiminarayanan16}. On a small and complex dataset, this complementarity is particularly valuable since no single feature extractor is likely to generalise well across the full range of source morphologies. In both architectures, one branch is identical to the simple CNN, and the other branches apply an SE block after each convolutional block. The two architectures differ in their input to the second branch.

The third model, the dual-branch convolutional squeeze network (DualCSN), is shown in Table~\ref{tab:DualCNNSqueezeNet}. It uses images as input for both branches. Due to the two-layer-longer linear network that merges the two branches, the DualCSN contains 388,530 parameters. The fourth model, the dual-branch scatter squeeze network (DualSSN), instead inputs the scattering coefficients to the SE-augmented branch. This substitution reduces the parameter count to 116,146, shown in Table~\ref{tab:DualScatterSqueezeNet}.

\subsection{Data and pre-processing} \label{sec:data_and_processing}

\subsubsection{The PSZ2 data}\label{sec:data}
We used images of galaxy clusters from the second LOFAR two-metre sky survey data release \citep[LoTSS-DR2;][]{shimwell22} which are cross-matched with the \textit{Planck} catalogue of Sunyaev--Zel'dovich (SZ) sources \citep[PSZ2;][]{planck15}. These images were analysed and processed by \cite{botteon22}, resulting in reference, \textit{uv}-tapered, and tapered and point-source-subtracted images:
\begin{itemize}
    \item Reference: Base images were produced with \textsc{WSClean} v2.8 using \cite{briggs95} weighting with a robust parameter of $-0.5$ \citep[Section 3.3;][]{botteon22}\footnote{\url{https://lofar-surveys.org/planck_dr2.html}}. The native beam size varies across the sample with a median of $13.5'' \times 8.4''$ (major $\times$ minor axis FWHM\footnote{Full width at half maximum}; range $5.8''$–$90.4''$), reflecting the heterogeneous declination coverage and observing conditions of the PSZ2 sample. After pre-processing, the reference images have a median pixel size of $7.2''$ (range $4.4''$–$10.4''$). These will be referred to as \emph{reference} images.
    \item \textit{uv}-tapered: For each cluster with known redshift, a Gaussian \textit{uv} taper commensurate with 25, 50, and \SI{100}{kpc} was applied on these visibilities, producing three different \textit{uv}-tapered versions, per image. The Gaussian \textit{uv}-tapered images on arbitrary scale \emph{X}$\,$kpc will be referred to as \emph{Tap. X$\,$kpc}.
    \item Point-source-subtracted: point-source-subtraction was done by removing the clean components obtained from imaging with a hard \textit{uv} cut corresponding to \SI{250}{kpc} for known cluster redshifts and 2722$\lambda$ for clusters without known redshifts. The same tapering procedure as above then produced three point-source-subtracted versions, per image. The point-source-subtracted images of type TapXkpc will be referred to as \emph{Tap.+Sub. X$\,$kpc}.
\end{itemize}
We used all three LoTSS-DR2 image types in this work because tapered image files show larger-scale emission more clearly and are expected to improve classification performance. Every image cutout is labelled as either diffuse emission (DE) or non-diffuse emission (NDE). The selected sample includes all images classified as radio halo (RH), radio relic (RR), or NDE, totalling 207 sources: 59 RH, 20 RR (of which 12 host both a halo and a relic, giving 67 DE sources in total), and 140 NDE. Candidate classifications are not included. 

\begin{figure}
    \centering
    \includegraphics[width=\columnwidth]{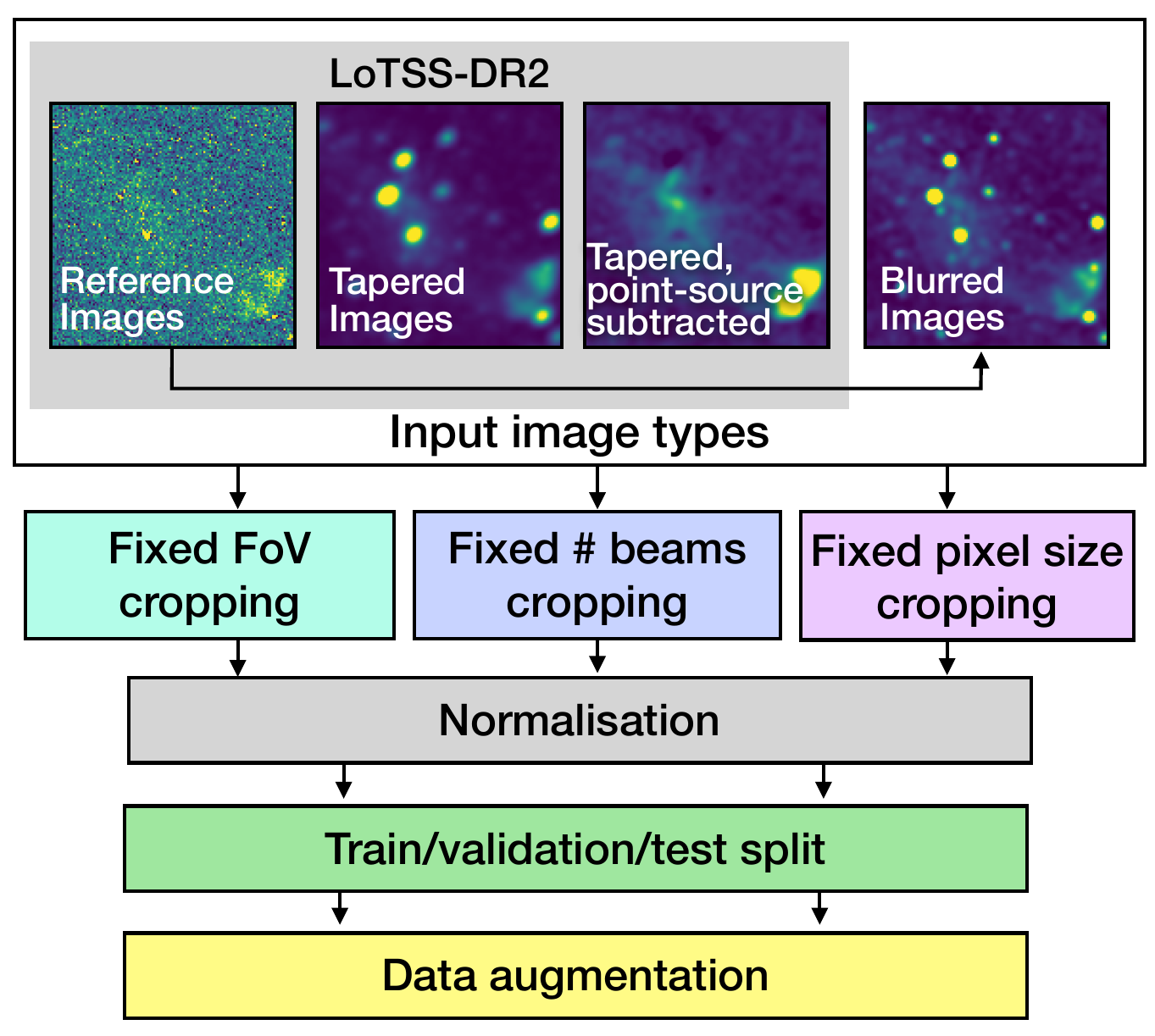}
    \caption{Different steps of the data pre-processing.}
    \label{fig:pre-processing}
\end{figure}

Every image is then processed as shown in Figure~\ref{fig:pre-processing}. Specifically, the three original data types from LoTSS-DR2 make four input image types after the reference images are blurred. All types except the reference type exist in the three resolution scales 25, 50, and \SI{100}{kpc}. We refer to each (type, scale) pair, as a version (ten in total: one reference plus three types times three scales). Each version can then be cropped in three parallel ways, making up 30 image variants that are then normalised, split, and augmented individually with rotations and horizontal flips. These steps are detailed below.

\subsubsection{Image blurring}
Diffuse cluster radio emission is characterised by large angular scales and low surface brightness, and is therefore often suppressed in high-resolution imaging. Tapering reduces the contribution of visibilities on long baselines, thereby decreasing sensitivity to small-scale structure and effectively lowering the image resolution. This preferentially enhances extended, low-surface-brightness emission, at the cost of angular resolution, relative to compact sources. 

However, $uv$-tapering requires access to the original visibilities, which may not be readily available for future large-scale surveys such as those conducted with the SKA due to data volume constraints and the use of standardised image products. To address this, we investigated image-domain blurring as a computationally efficient alternative that can be applied directly to restored maps, without reference to the visibilities or to any $uv$-tapered product. This is in contrast to \textit{uv}-tapering, which is applied to the visibilities prior to imaging: the taper weights modify the dirty image before cleaning, so that deconvolution is performed on an already-tapered dirty beam. Since \textit{uv}-tapering modifies the synthesised beam and thereby changes the cleaning process, the two approaches are not expected to produce identical results.

In the ideal case of perfect \textit{uv}-coverage, applying a Gaussian taper in $uv$-space corresponds to Gaussian blurring in the image plane, described in Appendix~\ref{app:blur}. We used this procedure to create another image type from the reference images:
\begin{itemize}
     \item Blurred: The reference images were convolved with a Gaussian in the image domain to mimic TapXkpc. These versions will be referred to as \emph{BlurXkpc}.
\end{itemize}

\begin{figure*}
    \centering
    \includegraphics[width=\textwidth]{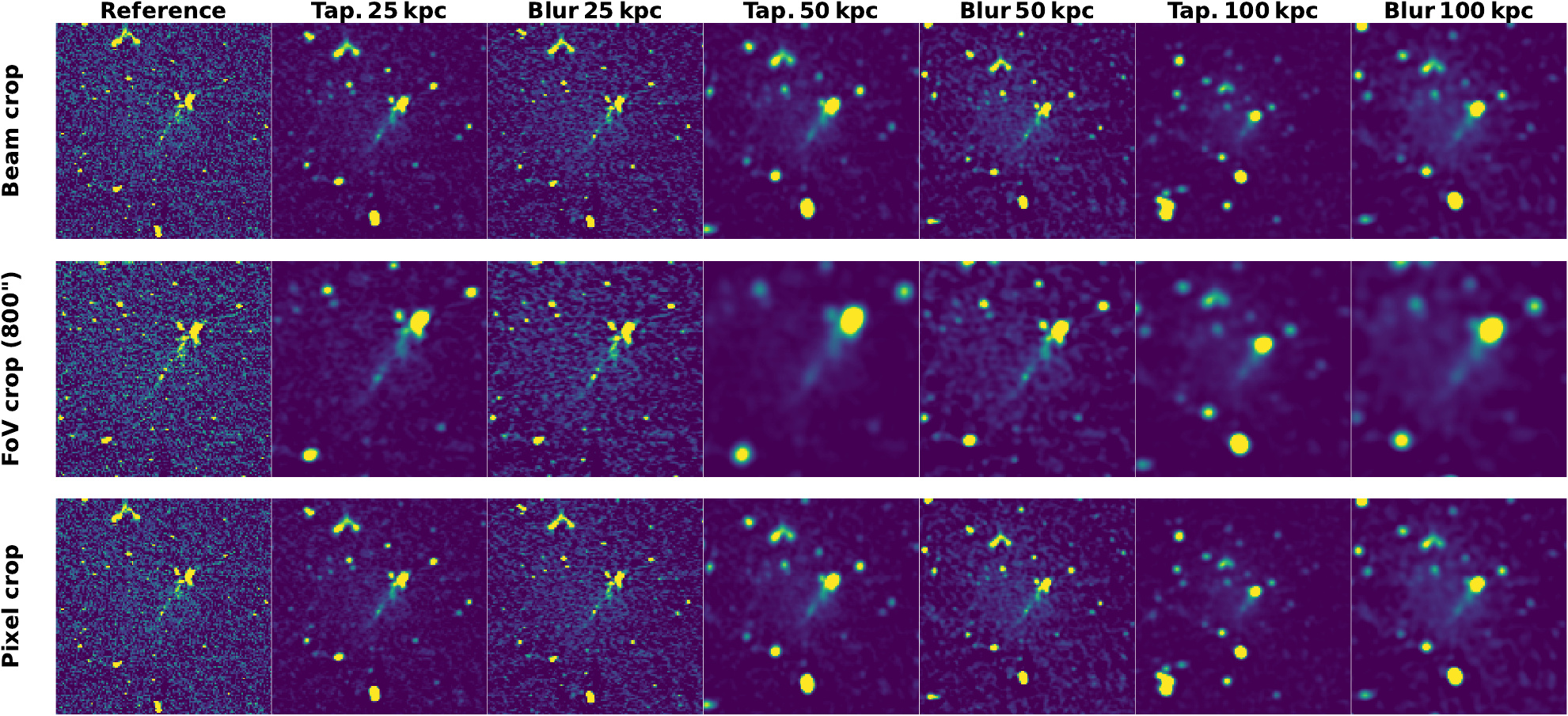}
    \caption{Examples of images processed under three cropping strategies (rows) across seven image versions (columns), for one randomly selected diffuse emission source. Each column presents a different image version: the reference image, followed by pairs of \textit{uv}-tapered \citep[produced by][]{botteon22} and image-blurred versions at 50$\,$kpc physical scales. The three rows correspond to: beam crop, where the FoV is set to a fixed multiple of the synthesised beam and equalised across versions; FoV crop, where a fixed angular FoV of 800'' is applied uniformly; and pixel crop, where the FoV is beam-proportional but cropped independently per version. The colour scale is the same per-image percentile normalisation as used in the classification pipeline.}
    \label{fig:gaussian-blurring}
\end{figure*}

\subsubsection{Image cropping}

Since the LoTSS images vary in pixel count, angular size, and physical scale, there was no obvious single cropping strategy that we could apply across the entire dataset. We therefore compared three different cropping strategies, summarised alongside the full processing pipeline in Table~\ref{tab:pipeline}. Since the three cropping strategies, as do the four image types and three tapering scales, likely accentuate different information, it is of interest to stack images with different type or scale of the same source as input to the classifiers. 

To align cluster centres and enable this multi-version stacking, each tapered and convolved image was first reprojected onto the pixel grid of the native-resolution image. This intermediate reprojection step ensured that all versions shared a common centre and orientation, but it did not fix a shared final pixel scale: after reprojection, each version was independently cropped to a version-specific field of view and resampled to $128\times128$ pixels, so the final pixel scale differs both across versions and across clusters. Pixels outside the image footprint were assigned NaN rather than zero, to prevent the classifier from treating absent coverage as true zero-flux. This introduced NaN-filled margins along the image edges whose width varies per cluster and per version.

To avoid cropping into these invalid regions, each cluster was cropped to its smallest NaN-free extent across all versions. The target field of view for each version is set in units of the local beam width, with the number of beams chosen so that all versions have the same mean field of view across the full sample. This equalisation is global rather than per-source: for any individual cluster the field of view still varies across versions (reflecting differences in beam size and NaN coverage), and across clusters it varies because the beam size differs. The field of view therefore remains close to the target $\theta$ for the majority of clusters but is not strictly constant.

\begin{itemize}
\item The pixel cropping method extracts the central $512\times512$ pixels of each image. This is a simple cropping strategy very commonly used for applications of machine learning in radio astronomy \citep[e.g.][]{Aniyan17, Gheller18, Maslej21, Mishra24}.
\item The field-of-view (FoV) cropping strategy extracts a fixed square region of $\theta=800''$ per side. By maintaining a constant FoV rather than a fixed pixel count, the strategy allows for better generalisation across diverse observations. 
Since different image resolutions correspond to different pixel scales, this method produces images with varying numbers of pixels that are later made equal in the universally applied downsample step. FoV cropping is conveniently done when simulated data is part of the training (see e.g. \citep{Stuardi21, Sanvitale25}).
\item The beam cropping strategy extracts a per-version (type and scale) fixed number of effective beam widths. In the case of reference images, the effective beam width equals that of the synthesised beam in the reference data. For other versions it is the target beam after blurring or $uv$-tapering, which means that beam sizes differ across target redshifts and tapering or blurring scales. The beam cropping strategy was chosen because the synthesised beam sets the fundamental resolution element of the image, and hence also the natural scale on which noise correlates spatially. By cropping to a fixed number of beam FWHMs, the noise correlation structure is approximately uniform across all images in the dataset, regardless of the native resolution or observing configuration of the underlying data. This is particularly important for \textit{uv}-tapered versions, where beam sizes vary substantially across images and tapering scales (see Figure~\ref{fig:gaussian-blurring}). For the beam cropping, the number of beams per image version was calibrated globally so that the mean angular field of view is approximately the same for all ten image versions. For each version, we scanned all images to find the minimum NaN-free pixel-to-beam ratio, and compute the resulting mean angular field of view. The version with the smallest mean field of view sets a global target, and the beam counts for all other versions are scaled down proportionally to match it. Each image is then cropped to its version-specific beam count times its own beam FWHM, capped at its NaN-free extent. The beam cropping strategy is, to our knowledge, novel in the context of radio astronomy classification. 
\end{itemize}

After applying the respective cropping strategy, all images were then downsampled to $128\times128$ pixels via bilinear interpolation \citep{torch-interpolate}. We computed the azimuthally averaged power spectra of the reference images before and after downsampling to $128\times128$ pixels for a representative sample of sources spanning the beam-size range of the dataset. The power spectra agree well at angular scales larger than the Nyquist limit of the downsampled images ($\sim$$14''$), confirming that the large-scale emission relevant for classification is preserved. Power at sub-beam scales is suppressed by the downsampling, but this regime carries no astrophysical information beyond the instrumental noise floor.

\begin{table}
  \centering
  \caption{Per-cluster processing pipeline.}
  \label{tab:pipeline}
  \setlength{\tabcolsep}{4pt}
  \begin{tabular}{clp{6.1cm}}
    \toprule
    Step & Operation & Description \\
    \midrule
    1 & Project
      & Reproject all versions onto a common pixel grid; pixels outside
        the image footprint are set to NaN. \\[4pt]
    2 & Crop
      & Centre-crop each version according to mode $M$:
        \smallskip
        \begin{tabular}[t]{lp{3.1cm}}
          \textit{Pixel} & Fixed $512\times512\,\mathrm{px}$ window. \\[2pt]
          \textit{FoV}   & Fixed $\theta=800''$, capped at the smallest
                           NaN-free extent. \\[2pt]
          \textit{Beam}  & $n_V^\star$ beam FWHMs, capped at
                           $\ell^{\rm NaN}_V$, with $n_V^\star$ equalising
                           the mean angular FoV across versions. \\
        \end{tabular} \\[4pt]
    3 & Resize
      & Bilinear interpolation to $128\times128\,\mathrm{px}$. \\
    \bottomrule
  \end{tabular}
  \tablefoot{Scales $\mathcal{S}=\{25,50,100\}$\,kpc. For beam cropping, $n_V^\star$ is a globally equalised beam multiplier (see text); $\ell^{\rm NaN}_V$ is the largest NaN-free centred square for version $V$ (for ${\rm Blur}_s$, taken from ${\rm Tap}_s$).}
\end{table}

We note that these cropping strategies are not exhaustive. For instance, one could crop the images to a constant physical size.

\subsubsection{Normalisation and augmentation} \label{sec:normalisation_and_augmentation}

After blurring and cropping, the pixel range of each individual image was clipped and linearly stretched between two hyperparameter percentiles, $p_{\text{lo}}=0.3$ and $p_{\text{hi}}=0.99$. Normalising pixel values to a fixed global range, rather than scaling each image individually, did not improve results. 
Clipped images were always arcsinh-stretched according to the mapping
\begin{equation}
I' \;=\; \frac{\operatorname{asinh}\!\left(\alpha\, I\right)}{\operatorname{asinh}(\alpha)}\,
\end{equation}
applied pixel-wise to the linearly rescaled pixel values $I \in [0,1]$ derived from the chosen percentile bounds. We used $\alpha=10$, which compresses bright, high-flux pixels while remaining approximately linear near the noise floor and is well defined for pixels with negative flux, unlike a pure $\log$ mapping. 

After blurring, cropping, and normalisation, the images were split into a fixed held-out test set (\SI{20}{\%}) and a training pool (\SI{80}{\%}), ensuring that the test dataset is identical across all runs. Within each training run, the training pool was further divided into train and validation subsets according to a ten-fold cross-validation. The training, validation, and test splits for each class are shown in Table~\ref{tab:dataset_composition}.

\begin{table}
\centering
\caption{Number of clusters per class and image type.}
\label{tab:dataset_composition}
\begin{tabular}{llcc}
\hline
Class & Split & Reference & Non-reference images \\
\hline
DE  & Train      & 46-51 & 42-49 \\
    & Validation & 3-8 & 1-8  \\
    & Test       & 13 & 17 \\
\hline
    & TOTAL      & 67 & 67 \\
\hline
NDE & Train      & 98-103 & 81-88 \\
    & Validation & 9-14 & 6-13 \\
    & Test       & 28 & 20 \\
\hline
    & TOTAL      & 140 & 114 \\
\hline
Total & --        & 207 & 181 \\
\hline
\end{tabular}
\tablefoot{DE: diffuse emission; NDE: no diffuse emission. The third column contains all catalogue clusters, the fourth counts only clusters with known redshift. All 26 sources without known redshift lack diffuse emission. The exact training-validation split varies due to the ten-fold cross-validation, which creates different amounts of training and validation images per split.}
\end{table}

After the training, validation, and test data split we applied data augmentation to each image, consisting of $n_{\text{rot}}=12$ different angular rotations and horizontal flipping. We did not use other augmentation techniques, as they are less likely to improve accuracy \citep{Aniyan17, Maslej21, ndungu23}. Both training and validation images were shuffled prior to each epoch.

\subsection{Loss function and training} \label{sec:training}

We used several regularisation strategies to train the networks. On small datasets, neural networks are prone to  overfitting: the model memorises training examples rather than learning generalisable features. This manifests as a growing gap between training  and validation loss. Regularisation counteracts this by constraining the hypothesis space, either by perturbing the input data, softening the supervision signal, or penalising large weights \citep{Srivastava14, Szegedy15, Zhang17, loshchilov17, Aniyan17}. Each strategy we employed targets a different aspect of this problem, and their combination is motivated by the fact that no single technique is sufficient when both the dataset noise and the label noise are substantial \citep{Muller19}. Label smoothing \citep{Szegedy15} with parameter $\varepsilon$ regularises the cross-entropy under uncertain or noisy labels. In the binary classification case, each one-hot target $y_i\!\in\!\{0,1\}$ indicates the true class of sample $i$, where $y_i=0$ denotes no diffuse emission (NDE) and $y_i=1$ denotes diffuse emission (DE). Label smoothing replaces each target with
\[
\tilde{y}_i \;=\; (1-\varepsilon)\,y_i \;+\; \frac{\varepsilon}{2}.
\]

We also implemented mix-up regularisation \citep{Zhang17} during training only. For each batch, we set a \SI{50}{\%} probability of creating linear combinations of the images, scattering coefficients, and labels, with the linear parameter, $\lambda_{\text{mu}}$, being drawn from a beta distribution:
\begin{equation}
    \lambda_{\text{mu}} \sim f(x;\alpha,\beta) = \frac{x^{\alpha-1}(1-x)^{\beta-1}}{\int_0^1 t^{\alpha-1}(1-t)^{\beta-1}\,\mathrm{d}t},
\end{equation}
where $\alpha = \beta = 0.4$ in this work. The mixed samples and loss were then computed as
\begin{align}
    \tilde{\mathbf{x}} &= \lambda_{\text{mu}}\, \mathbf{x}_i + (1 - \lambda_{\text{mu}})\, \mathbf{x}_j, \\
    \mathcal{L}_{\text{mix}} &= \lambda_{\text{mu}}\, \mathcal{L}(\tilde{y}_i) + (1-\lambda_{\text{mu}})\, \mathcal{L}(\tilde{y}_j),
\end{align}
where $\mathbf{x}$ denotes either the image or scattering coefficient input, and $(i, j)$ were randomly paired samples within the batch. $\mathcal{L}$ is the cross-entropy loss
\begin{equation}
    \mathcal{L} = -\frac{1}{N}\sum_{n=1}^{N} w_{y_n} \sum_{c=1}^{C} \tilde{y}_{n,c} \log \frac{\exp(\hat{y}_{n,c})}{\sum_{c'=1}^{C} \exp(\hat{y}_{n,c'})},
\end{equation}
where $N$ is the number of samples in the batch, $C=2$ is the number of classes, $w_{y_n}$ is the weight assigned to the smoothed true class $\tilde{y}_n$ of sample $n$, $\tilde{y}_{n,c}$ is the smoothed label for sample $n$ and class $c$, and $\hat{y}_{n,c}$ is the predicted score (logit) for sample $n$ and class $c$. 

We trained the classifier with the AdamW optimiser \citep{loshchilov17} and a batch size of 16. Small batch sizes introduce stochastic gradient noise, which acts as an implicit regulariser and helps the optimiser escape sharp minima that tend to generalise poorly \citep{Keskar16}. This is particularly relevant here given the limited number of training samples. For stability on the small and complex dataset, a smaller learning rate of $4\times10^{-5}$ and a larger L$_2$ regularisation of $\lambda_{L2}=0.1$ were chosen. $L_2$ regularisation adds a term $\lambda_{L2}\|w\|_2^2$ to the loss function, penalising large weight magnitudes and discouraging the network from relying too heavily on any single feature \citep{krogh91}. All experiments were conducted on the Alps supercomputer at the Swiss National Supercomputing Centre (CSCS), an HPE Cray EX system. Each configuration of dataset version and crop mode was run for ten different training-validation splits, repeated three times each, for a total of 30 experiments per configuration.

For each run, four standard scores were computed: accuracy, recall, precision, and \(F_1\). We use acronyms where \(\mathrm{T}\) and \(\mathrm{F}\) denote true and false, and \(\mathrm{P}\) and \(\mathrm{N}\) denote positive and negative, respectively. Throughout, DE is taken as the positive class and NDE as the negative class. We evaluated the following metrics:
\begin{itemize}
    \item Accuracy \(=(\mathrm{TP}+\mathrm{TN})/(\mathrm{TP}+\mathrm{TN}+\mathrm{FP}+\mathrm{FN})\), the fraction of correct classifications. 
    \item Recall \(=\mathrm{TP}/(\mathrm{TP}+\mathrm{FN})\), the completeness of DE recovery. 
    \item Precision \(=\mathrm{TP}/(\mathrm{TP}+\mathrm{FP})\), the purity of predicted DE.
    \item \(F_1 = 2\,\mathrm{TP}/(2\,\mathrm{TP}+\mathrm{FP}+\mathrm{FN})\), the trade-off between recall and precision.  
\end{itemize} 

\section{Results} \label{sec:results}

\FloatBarrier
\subsection{Reference data}

\begin{figure}
    \centering
    \includegraphics[width=\hsize]{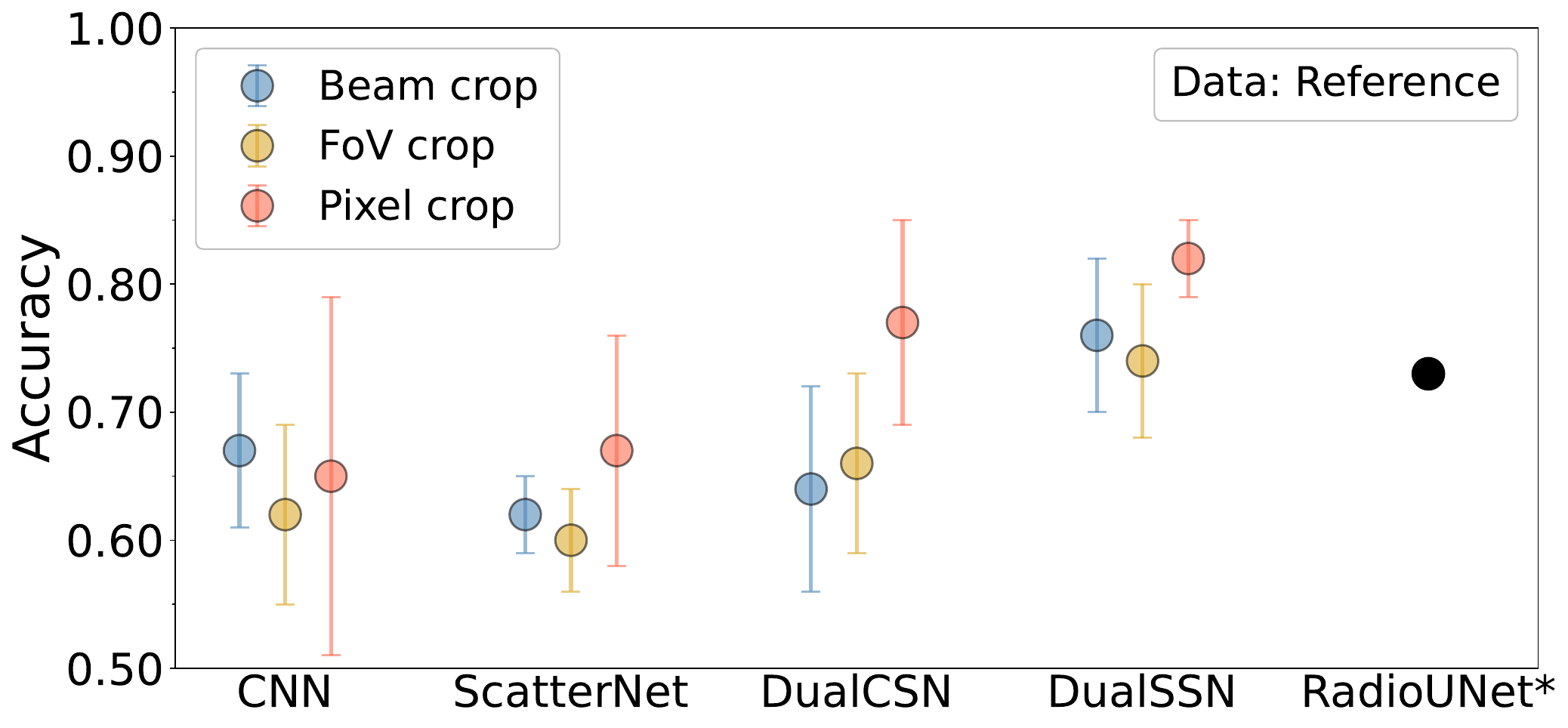}
    \caption{Diffuse emission classification accuracy on the reference images (native $\sim$6\arcsec\ resolution, no \textit{uv}-tapering, point-source-subtraction, or blurring) for all four classifiers and three cropping strategies. Each point shows the mean over 30 runs; error bars show the standard deviation. The RadioUNet point shows the accuracy reported by \protect\cite{Stuardi24} on the LoTSS-DR2/PSZ2 sample using 20\arcsec\ resolution archive images, included as a rough contextual benchmark.}
    \label{fig:accuracy_reference_data}
\end{figure}

First, we evaluated the classification performance for all three image cropping strategies on the reference images, images without any \textit{uv}-tapering, point-source-subtraction, or blurring.
The results for different neural network architectures are shown in Figure~\ref{fig:accuracy_reference_data}, and also includes the accuracy of \cite{Stuardi24} shown as a single black dot. Their result of 0.73 was obtained on the same LoTSS-DR2/PSZ2 sample but used the standard 20\arcsec\ low-resolution archive images, whereas the reference images used here are the post-processed $\sim$6\arcsec\ images from \cite{botteon22}. The two performances are therefore not directly comparable, and the RadioUNet accuracy is included only as a rough contextual benchmark.

As further detailed in Appendix~\ref{sec:additional_results}, the DualSSN consistently achieves the highest accuracy, $F_1$, and area under curve (AUC) for the receiver operating characteristics (ROC) in all three cropping settings, on the reference data. It is also outperforming RadioUNet for all three cropping settings. The detailed metrics for beam-cropped data in Table~\ref{tab:reference_beamcrop} are reported here since beam cropping is the best-performing strategy overall (see Section~\ref{sec:blurred_tapered_results}), despite pixel cropping achieving higher accuracy on reference data specifically. The table shows that while the DualCSN achieves the highest recall ($0.79 \pm 0.16$), it does so at the cost of precision ($0.60 \pm 0.10$), resulting in a lower $F_1$ than the DualSSN. The ScatterNet is the worst-performing classifier on the reference data. It also has the smallest standard deviations for the beam- and FoV-cropped data.

Aside from the CNN, we obtained the highest classification accuracy with pixel-cropped data, while FoV-cropped data results in the worst performance. For the reference images, all types and scales share a uniform and high native resolution, so beam cropping provides no noise-normalisation advantage over pixel cropping. The superior performance of the pixel cropping likely reflects the absence of resampling artefacts introduced by the bilinear interpolation step required in beam and FoV cropping. The FoV crop performs worst, which may reflect the fact that a fixed angular window captures physically very different regions at different redshifts.

\begin{figure}
  \centering
  \includegraphics[width=\hsize]{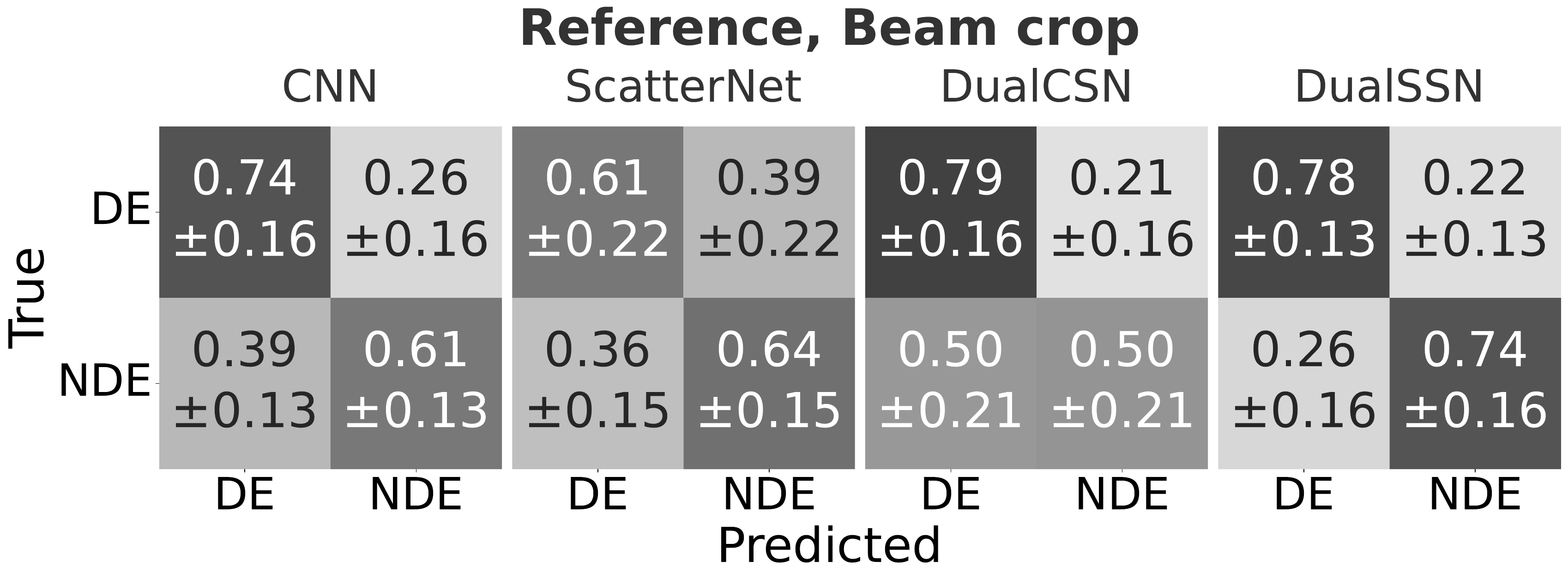}
  \caption{Confusion matrices for the four different classifiers on beam-cropped reference data. Standard deviations are calculated from the 30 individual runs that vary random seed and cross-validation fold.}
\label{fig:cm_reference}
\end{figure}

Figure~\ref{fig:cm_reference} shows the corresponding confusion matrices. The confusion matrix of DualSSN shows a more balanced error distribution between false positives and false negatives than the other models. 

\begin{table}
\centering
\caption{Classification performance on reference images using beam-normalised cropping.}
\label{tab:reference_beamcrop}
\small
\begin{tabular}{lcccc}
\hline
Metric & CNN & ScatterNet & DualCSN & DualSSN \\
\hline
Acc.  & $0.67 \pm 0.06$ & $0.62 \pm 0.03$ & $0.64 \pm 0.08$ & $\mathbf{0.76 \pm 0.06}$ \\
Prec. & $0.60 \pm 0.12$ & $0.59 \pm 0.03$ & $0.60 \pm 0.10$ & $\mathbf{0.74 \pm 0.11}$ \\
Rec.  & $0.74 \pm 0.16$ & $0.61 \pm 0.22$ & $\mathbf{0.79 \pm 0.16}$ & $0.78 \pm 0.13$ \\
$F_1$ & $0.66 \pm 0.13$ & $0.58 \pm 0.12$ & $0.66 \pm 0.13$ & $\mathbf{0.75 \pm 0.05}$ \\
AUC   & $0.74 \pm 0.07$ & $0.67 \pm 0.03$ & $0.74 \pm 0.09$ & $\mathbf{0.85 \pm 0.05}$ \\
\hline
\end{tabular}
\tablefoot{Averaged over 30 runs. Metrics reported as mean $\pm$ standard deviation. Acc: accuracy. Prec: precision. Rec: recall. Best result per row in \textbf{bold}.}
\end{table}

\subsection{Blurred, tapered and point-source-subtracted data} \label{sec:blurred_tapered_results}

\begin{figure}
    \centering
    \includegraphics[width=\hsize]{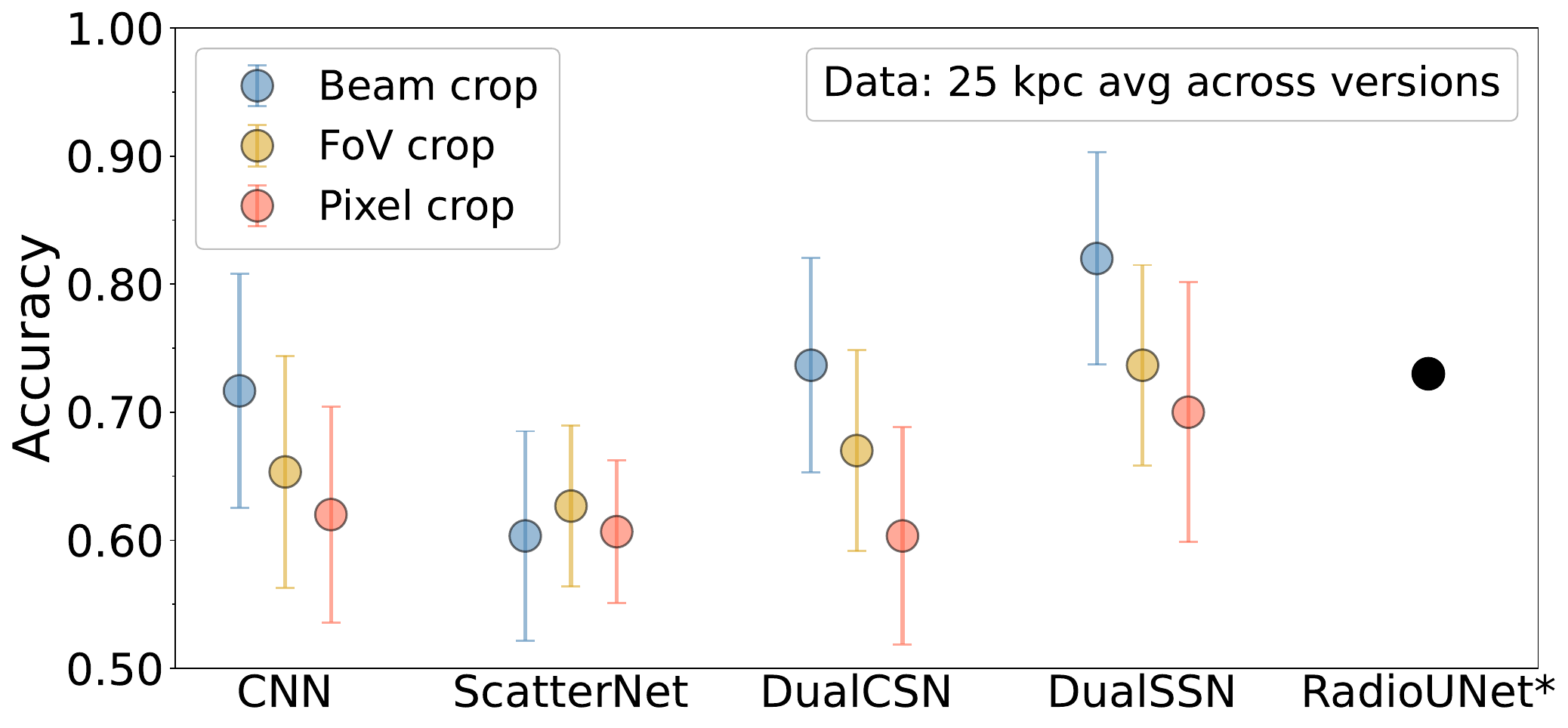}
    \caption{Classification accuracy for all four classifiers and three cropping strategies, averaged over the three 25$\,$kpc dataset versions (blurred, \textit{uv}-tapered, and \textit{uv}-tapered and point-source-subtracted). Each point shows the mean over 90 runs (30 per version); error bars show the standard deviation. Beam cropping consistently outperforms FoV and pixel cropping for all classifiers except ScatterNet. The RadioUNet point is from \protect\cite{Stuardi24}, included as a rough contextual benchmark.}
    \label{fig:classification_accuracy_25kpc}
\end{figure}

\begin{figure}
    \centering
    \includegraphics[width=\hsize]{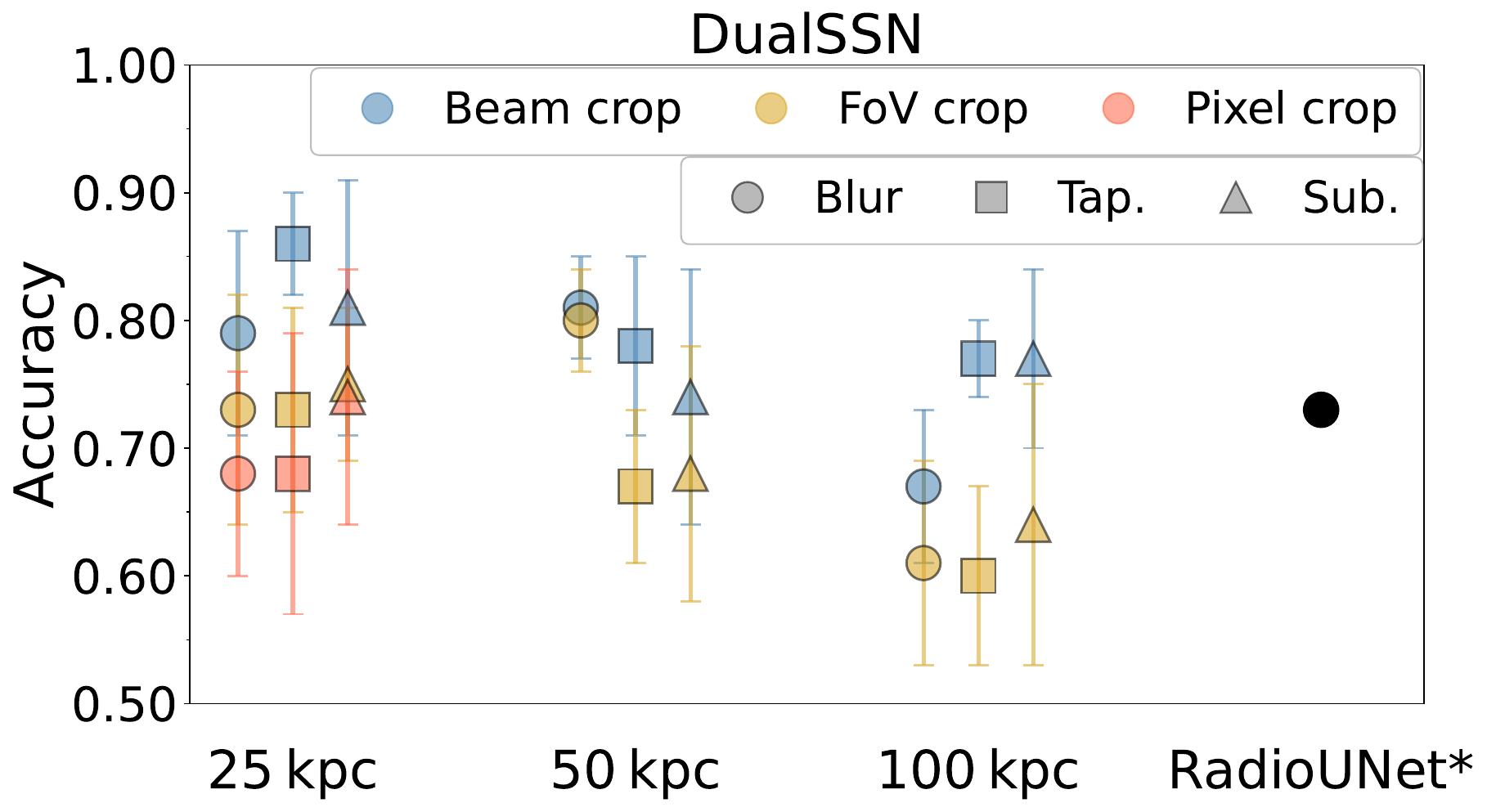}
    \caption{DualSSN classification accuracy across tapering/blurring scales ($X$ = 25, 50, and 100\,kpc), cropping strategies (beam, FoV, pixel), and all three dataset types (blurred, \textit{uv}-tapered, and point-source-subtracted). Each point shows the mean and standard deviations over 30 runs. Performance degrades with increasing scale for all cropping strategies, and beam cropping consistently outperforms FoV and pixel cropping. The RadioUNet point is from \protect\cite{Stuardi24}, included as a rough contextual benchmark.}
    \label{fig:dualssn_accuracy_scatter}
\end{figure}

Figure~\ref{fig:classification_accuracy_25kpc} presents the classification accuracy across all classifiers and crop modes, averaged over blurred, \textit{uv}-tapered, and \textit{uv}-tapered and point-source-subtracted data using $X=$\SI{25}{kpc}. We found that using beam cropping consistently outperforms FoV and pixel cropping, with DualSSN beam cropping achieving the highest accuracy overall. This ranking holds across classifiers, with the exception of the ScatterNet. 
Unlike the reference data, the 25\,kpc versions have substantially varying effective beam sizes across cluster images and version scales, meaning that pixel and FoV cropping capture an inconsistent number of beams per image. Beam cropping removes this variation, rendering the noise texture more uniform across the dataset and making morphological differences between DE and NDE images more discriminative. The ScatterNet is an exception to this ranking, possibly because its fixed wavelet kernels are less sensitive to the noise-scale variation that disadvantages the other classifiers under pixel and FoV cropping.
Given the substantial accuracy gap between beam and pixel cropping on these data, no classifiers were trained on pixel-cropped data using scales of 50\,kpc and 100\,kpc. The remainder of this chapter, with the exception of Figure~\ref{fig:dualssn_accuracy_scatter}, excludes pixel cropping.

Figure~\ref{fig:dualssn_accuracy_scatter} shows the accuracy of the best performing classifier, DualSSN, across all three tapering scales and cropping strategies. As for the CNN and DualCSN on data using scale $X=25$\,kpc, beam cropping consistently outperforms FoV and pixel cropping at all scales. We also see that performance degrades with increasing blurring or tapering scale. That is likely because coarser resolution blends compact sources more heavily with any underlying diffuse emission, which reduces the contrast between DE and NDE images.
Of the beam-cropped data, the different image processing types (blurred, tapered, or tapered and point-source-subtracted) have comparable accuracy, though tapered data performs marginally better than or equal to the point-source-subtracted equivalent. This is consistent with point-source-subtraction being an imperfect process: residuals from over- or under-subtracted compact sources near cluster centres introduce additional artefacts that may obscure the diffuse emission signal rather than clarify it. The tapered images retain compact sources in a smoothed, predictable form that the classifier can learn to discount.

The best-performing configuration is DualSSN with beam-cropped \textit{uv}-tapered 25\,kpc data. For this model and dataset version, the beam cropping outperforms FoV and pixel cropping on all five metrics, with an accuracy of $0.86 \pm 0.04$, averaged over all 30 runs. As shown in Table~\ref{tab:ensemble_top_n}, the best individual run achieves an accuracy of 0.92. Combining the five highest-accuracy runs via soft-voting (top-5 ensemble) improves this to 0.94. This is expected: each run is trained with a different random seed, so the models make different mistakes. When their predicted probabilities are averaged, errors that appear in only one or two runs are outvoted by the correct predictions of the others, yielding a more reliable result than any single run alone. Using all 30 runs (top-30) instead lowers performance, since lower-quality runs dilute the ensemble.

\begin{table}
  \centering
  \caption{Soft-voting ensemble performance on the fixed test set.}
  \label{tab:ensemble_top_n}
  \begin{tabular}{lcccc}
    \toprule
    Ensemble & Accuracy & Recall & Precision & F1  \\
    \midrule
    Top-1 & 0.92 & 0.90 & 0.93 & 0.91\\
    Top-5 & \textbf{0.94} & \textbf{0.91} & \textbf{0.95} & \textbf{0.93} \\
    Top-30 & 0.90 & 0.89 & 0.89 & 0.89 \\
    \bottomrule
  \end{tabular}
  \tablefoot{DualSSN, \SI{25}{kpc} \textit{uv}-tapered data with beam cropping. Runs are ranked by individual test accuracy; top-$N$ means the $N$ highest-accuracy runs are combined by averaging their predicted probabilities. Best result per column in \textbf{bold}.}
\end{table}

\begin{figure*}
    \centering
    \includegraphics[width=\hsize]
    {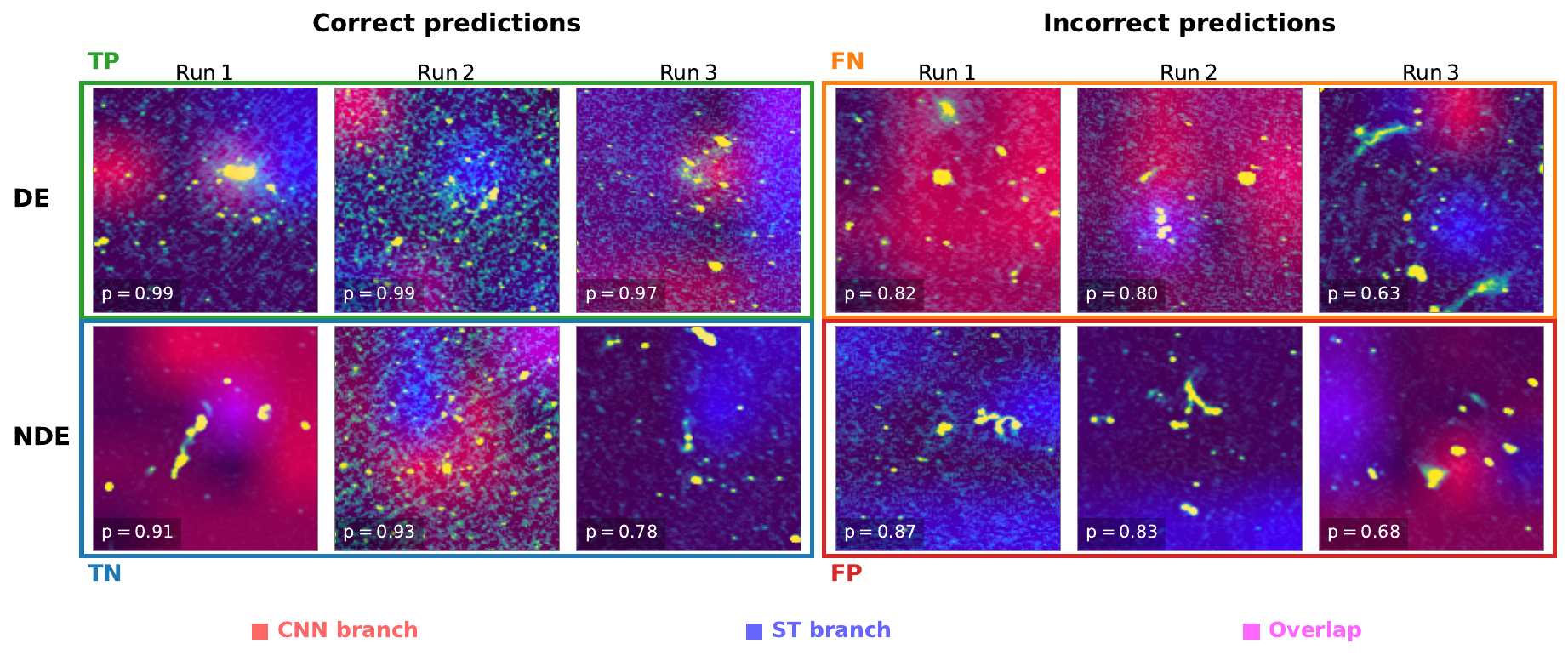}
    \caption{Six correctly classified images (left) and six erroneously classified images (right) from the DualSSN on \SI{25}{kpc} \textit{uv}-tapered beam-cropped test data, selected across three training runs. The top row shows diffuse-emission images (true positives, left; false negatives, right); the bottom row shows no-diffuse-emission images (true negatives, left; false positives, right). Overlaid in red and blue are the Grad-CAM attention maps \citep{selvaraju17} for the CNN and scattering transform branches respectively, with their overlap shown in magenta; warmer colours indicate regions that contributed more strongly to the DE prediction. The assigned DE probability for that run is printed in the bottom left of each panel.}
    \label{fig:best_example_predictions}
\end{figure*}

Figure~\ref{fig:best_example_predictions} displays twelve predictions from the DualSSN on \SI{25}{kpc} \textit{uv}-tapered beam-cropped test data. To visualise which regions of the input drive the classifier decision, we overlayed Grad-CAM attention maps \citep{selvaraju17}, which use the gradients of the predicted class flowing into the final convolutional layer to highlight the image regions most influential to that prediction. In red and blue we show the Grad-CAM attention maps of the CNN and ST branches, respectively. The classifier appears to perform better on (1) images where the diffuse emission is more extended and (2) central radio halos. The correctly classified NDE images are dominated by a large, bright radio-galaxy-like source. For instance, run~3 of the false negatives displays a clear double relic. Run~1 and run~2 of the false positives contain more irregularly shaped bright emission and are both high confidence scores. Further, all three examples of false negatives contain a dominant compact source closer to the centre than the diffuse emission. That is not the case for the correctly predicted DE images. Run~1 of the false negatives has a high confidence score of 0.82, with diffuse emission near the upper edge of the image, directly above a larger compact source. The model is least confidently wrong in false negative run~3 and false positive run~3, both of which contain a small amount of more centrally located diffuse emission.
The CNN branch activates broadly over large-scale, low-surface-brightness structure near the image centre, in the proximity of sources that are both irregular and bright, and produces weak activation for images it classifies as NDE. The scattering branch maintains a more comparable level of activation across all twelve examples and tends to attend to the image periphery, complementing the central focus of the CNN branch. The overlap between the two branches is concentrated and source-centred for true positives, broad and diffuse for false positives, and weak for true and false negatives. 

\par

\begin{figure*}
    \centering
    \includegraphics[width=\hsize]{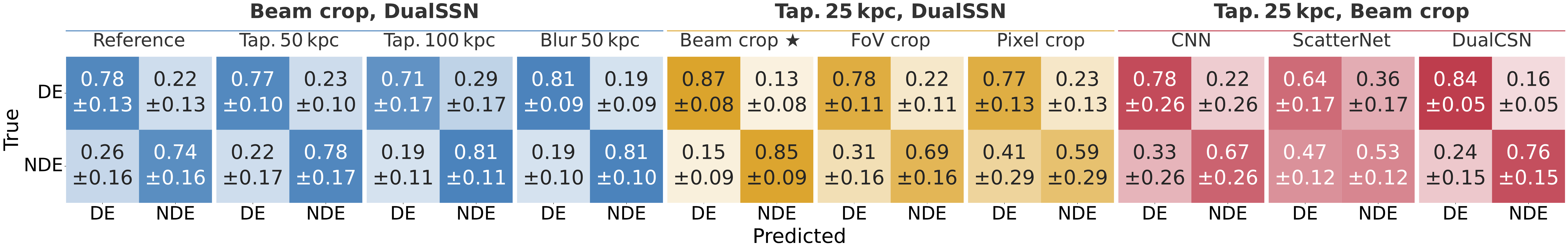}
    \caption{Confusion matrices comparing image versions using DualSSN on beam-cropped data (four matrices to the left), cropping modes for \SI{25}{kpc} \textit{uv}-tapered data with DualSSN (three matrices in the middle), and classifiers on \SI{25}{kpc} \textit{uv}-tapered data beam-cropped data (three matrices in the middle right). The star marks the best classifier. Standard deviations are derived from the 30 runs behind the mean.}
    \label{fig:confusion_matrices}
\end{figure*}

\begin{figure*}
    \centering
    \includegraphics[width=\textwidth]{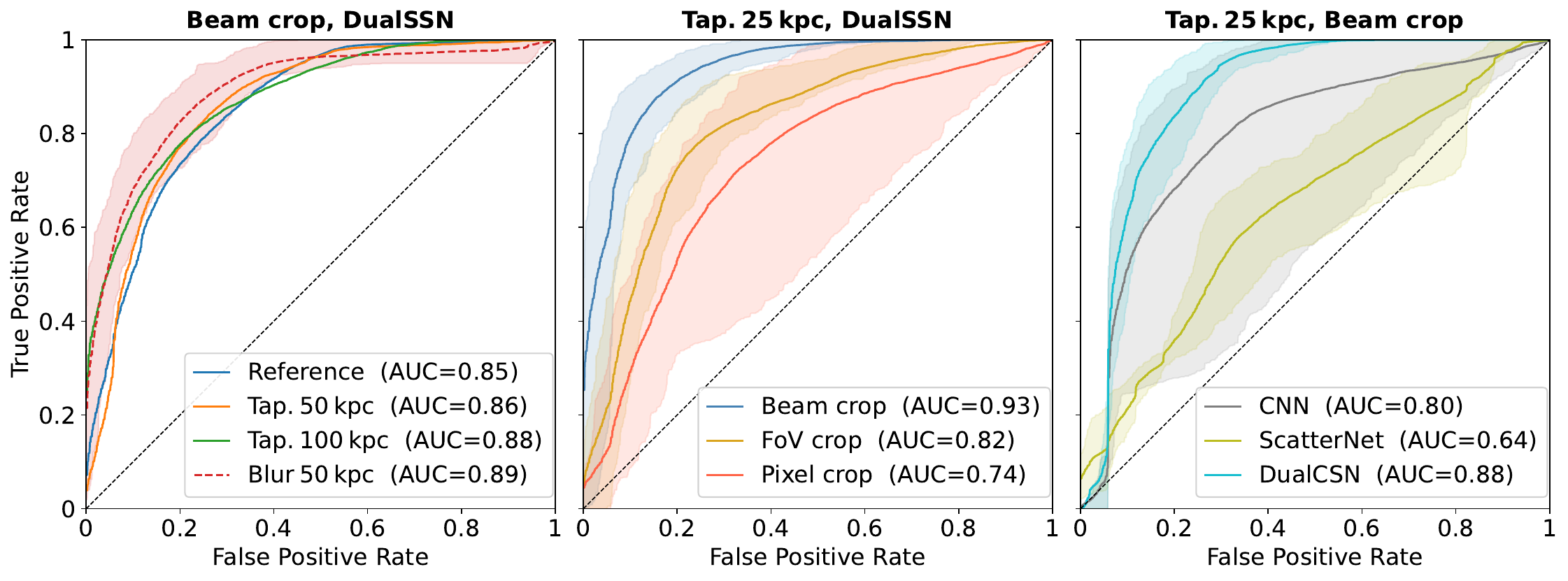}
    \caption{ROC curves comparing (left) dataset versions using DualSSN on beam-cropped data, (middle) cropping modes for \SI{25}{kpc} \textit{uv}-tapered data with DualSSN, and (right) classifiers on \SI{25}{kpc} \textit{uv}-tapered data beam-cropped data. Standard deviation regions are coloured. For the left plot, only the standard deviations for Blur 50$\,$kpc is shown, for the purpose of clarity.}
\label{fig:roc_curves}
\end{figure*}

Confusion matrices obtained by varying each hyperparameter individually around the best-performing configuration are shown in Figure~\ref{fig:confusion_matrices}. The false positive rate is substantially lower for the beam-crop than for the other modes. The ScatterNet and CNN have high rates of false positives. This confusion matrix imbalance is the smallest for beam-cropped data. The ROC curves in Figure~\ref{fig:roc_curves} reinforce this, with the beam-crop AUC exceeding the FoV and pixel crops by 0.11 and 0.19 respectively. The two figures also include the results of beam-cropped Blur 50\,kpc with DualSSN, which is the best performing blurred dataset classifier. This classifier has a perfectly symmetric confusion matrix, and a more asymmetric ROC curve, where it is very confidently wrong about $\sim5\,\%$ of the false positives.

\subsection{Stacked data}
Blurring and \textit{uv}-tapering enhance extended structure, but also increase the angular extent of bright point sources. Instead of only using one image version, classification performance may improve by using several versions at once. We extended the two-dimensional image input to three-dimensional cubes by stacking different image versions of the target. Table~\ref{tab:stacking_dualssn} summarises the DualSSN stacking performance across input combinations using beam cropping, where the 50\,kpc scale achieves the highest scores across most metrics.

We do not find that stacking improves the classification performance. This is likely because the blurred and reference versions of the same images are highly correlated: they encode the same underlying morphology, differing primarily in resolution and noise level rather than in independent discriminative content. On a small dataset, the additional input channels increase model complexity without a commensurate gain in information. This contrasts with the motivation for multi-branch architectures, where the two branches process fundamentally different feature representations, rather than two differently smoothed versions of the same image.

\begin{table}
\centering
\caption{DualSSN stacking performance across input combinations using beam-normalised cropping.}
\label{tab:stacking_dualssn}
\setlength{\tabcolsep}{4pt}
\begin{tabular}{lcccc}
\hline
Metric & 25\,kpc & 50\,kpc & 100\,kpc & All \\
\hline
Acc.  & $0.78 \pm 0.08$ & $\mathbf{0.79 \pm 0.05}$ & $0.73 \pm 0.07$ & $0.74 \pm 0.08$ \\
Prec. & $\mathbf{0.80 \pm 0.14}$ & $0.78 \pm 0.10$ & $0.73 \pm 0.12$ & $0.73 \pm 0.12$ \\
Rec.  & $0.77 \pm 0.14$ & $\mathbf{0.79 \pm 0.08}$ & $0.74 \pm 0.13$ & $0.75 \pm 0.13$ \\
$F_1$ & $0.76 \pm 0.07$ & $\mathbf{0.78 \pm 0.04}$ & $0.72 \pm 0.04$ & $0.73 \pm 0.08$ \\
AUC   & $\mathbf{0.88 \pm 0.05}$ & $0.87 \pm 0.05$ & $0.82 \pm 0.04$ & $0.83 \pm 0.07$ \\
\hline
\end{tabular}
\tablefoot{Each column corresponds to the reference image stacked with a single blurred version at the indicated physical scale, or all four versions stacked together (All). See Table~\ref{tab:reference_beamcrop} for metric definitions and averaging details.}
\end{table}

\subsection{Size and noise stability} \label{sec:size_and_noise}

\begin{figure}
    \centering
    \begin{subfigure}{\hsize}
        \centering
        \includegraphics[width=\hsize]{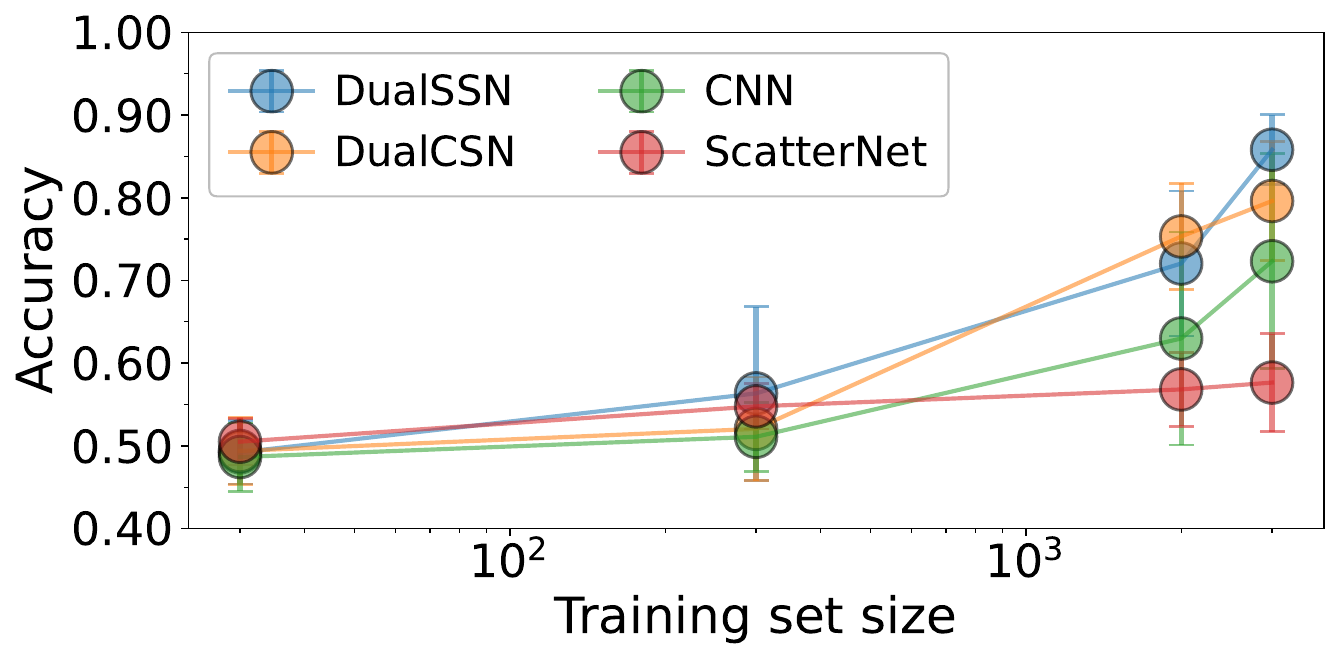}
        \caption{Accuracy as a function of training dataset size.}
        \label{fig:learning_curve}
    \end{subfigure}
    \vspace{0.5em}
    \begin{subfigure}{\hsize}
        \centering
        \includegraphics[width=\hsize]{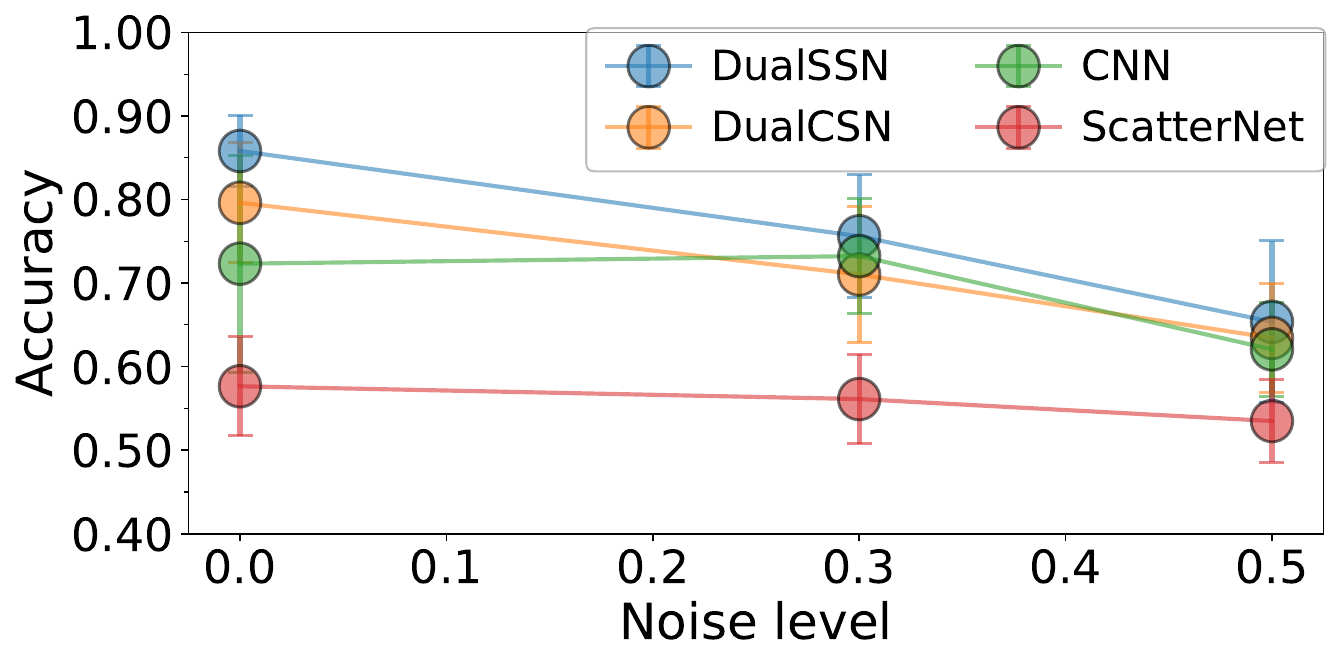}
        \caption{Accuracy as a function of injected noise level.}
        \label{fig:noise_sweep}
    \end{subfigure}
    \caption{Size and noise stability of the four classifier architectures on beam-cropped 25$\,$kpc \textit{uv}-tapered data. Error bars represent the standard deviation over 30 runs.}
    \label{fig:stability}
\end{figure}

To assess the dependence of classifier performance on training set size, we randomly subsampled the training and validation sets to 0.01, 0.1, 0.5, and 1 of the full training set, while keeping the test set fixed and identical across all configurations to ensure that reported accuracies are directly comparable. The results are displayed in Figure~\ref{fig:learning_curve}. The ST appears to improve performance on smaller datasets, because ScatterNet is the best model on small datasets and worst on the largest datasets. The DualSSN, which uses the ST as well as a CNN branch, performs comparatively well on all dataset sizes. 
Figure~\ref{fig:noise_sweep} compares the performance of the four classifiers when trained and evaluated on noisier data. We applied a noise level parameter $n_l \in \{0, 0.3, 0.5\}$ that linearly interpolates between the (normalised) clean image $\mathbf{x}$ and pure Gaussian noise,
\begin{equation}
    \mathbf{x}_{n_l} = (1-n_l)\,\mathbf{x} + n_l\,\boldsymbol{\varepsilon}, \qquad \boldsymbol{\varepsilon} \sim \mathcal{N}(0,1),
\end{equation}
applied after normalisation, such that $n_l=0$ recovers the clean image. Each classifier is trained and evaluated at the same noise level. No classifier appears especially robust against noisier datasets.

\section{Discussion} \label{sec:discussion}

Accuracy variation: the high run-to-run variance in all results is a consequence of the small sample size, with validation folds containing as few as one DE cluster. Further, the error bars may be underestimated, as the test set is fixed across runs. This makes the choice of architecture and pre-processing particularly consequential: dual-branch architectures, fixed feature extractors and physics-motivated cropping offer partial compensation, as discussed below.

Reference, blurred, or \textit{uv}-tapered:
reference data generally result in worse performance than blurred and tapered data. This is not surprising as the defining marks of diffuse emission are its faintness and extension which are enhanced by \textit{uv}-tapering or blurring. One might expect the convolutional kernels of a neural network to replicate the blurring effect in the feature extraction layers.
However, the network only operates on images that are already cropped and downsampled, without sufficient resolution to reproduce the blurring effect internally. Furthermore, on a dataset of this size, the network is unlikely to reliably learn a smoothing operation as an implicit pre-processing step when it must simultaneously learn to classify. By feeding the network $uv$-tapered or blurred images directly, the relevant diffuse emission features are made explicit in the input, reducing the burden on the classifier.

The comparable performance between uv-processed and blurred data means that image-blurring is a viable alternative for classification tasks of complex radio data where no \textit{uv} manipulation is possible. However, as shown in Figure~\ref{fig:gaussian-blurring}, the image-blurring pipeline produces noisier images. It is possible that this effect can be reduced by a longer optimisation of the percentile normalisation coefficients or another downsampling method.

Image cropping strategy:
the beam cropping tends to outperform FoV and pixel cropping across most dataset versions and classifiers.
This is likely because cropping to a fixed number of beams standardises the noise correlation scale across images. This makes the morphological differences between DE and NDE images the dominant source of variation in the dataset rather than differences in noise texture.
This effect is most pronounced for the \textit{uv}-tapered versions, where effective beam sizes vary substantially across cluster images and processing scales (see Figure~\ref{fig:scatter_taper}), meaning that FoV and pixel cropping capture very different numbers of beams per image. The FoV crop performs worst overall, which may reflect the fact that a fixed angular window captures physically very different regions at different redshifts.
For the reference data, the pixel cropping yields the highest accuracy (see Figure~\ref{fig:accuracy_reference_data}), likely because it avoids resampling artefacts introduced by the bilinear interpolation used in beam and FoV cropping.
However, for the blurred and \textit{uv}-tapered versions, this advantage is offset by a different problem: because pixel cropping fixes the number of pixels rather than the number of beams, and beam size varies across images within the same version (see Figure~\ref{fig:scatter_taper}), the cropped images no longer correspond to a consistent physical or beam-relative scale. This reintroduces the scale inconsistency that beam cropping is designed to avoid.

Best neural network architecture:
of the four classifiers considered in this work, the DualSSN provides the best performance, especially for \textit{uv}-tapered and beam-cropped data. This is consistent with a complementarity between the fixed scattering transform, which provides stable generalisation, and the convolutional branch, which 
adapts to finer discriminative features.
Since the convolutional kernels are learned from a small training set, they risk converging to dataset-specific features rather than generalisable representations of diffuse emission. Thus, the general filters of the ST might lead to lower variance between dataset folds. As shown in Appendix~\ref{sec:additional_results}, these two models are also the most stable. The ScatterNet is also the only model that performs better for the more noisy blurred data compared to the \textit{uv}-tapered versions. For blurred images using $X=$100\,kpc, the ScatterNet is the best of all four networks.  When datasets are both small and complex, stable feature extraction and performance can be crucial for avoiding overfitting.
The consistently superior performance of the DualSSN therefore suggests that combining a fixed wavelet-based feature extractor with a learned convolutional branch is a more robust strategy than purely learned architectures when both the dataset and the morphological complexity of the sources are limited.

Compared to RadioUNet, the DualSSN also achieves higher accuracy and precision but lower recall. \cite{Stuardi24} obtained accuracy, precision, and recall of 0.73, 0.70, and 0.89 respectively on reference data, using a U-Net fine-tuned on a subset of the same LoTSS-DR2 sample and evaluated on a held-out test set of 94 clusters. The higher recall of RadioUNet is partly a consequence of its segmentation-based detection strategy, which optimises a spatial probability threshold rather than a global image-level label, making it more sensitive to partial or marginal detections. Furthermore, \cite{Stuardi24} trained on a large set of synthetic observations and fine-tuned on real data, whereas the classifiers in this work are trained exclusively on the small labelled sample of real observations. Direct comparison is therefore limited, as the two methods differ in both architecture and training regime.

Generalisability:
A common issue with machine learning in astronomy is that models are highly dataset-dependent. A trained model often fails to generalise to unseen examples, and an architecture that is suitable for one dataset is not suitable for another. We deliberately avoided extensive hyperparameter tuning, as parameters optimised for one configuration are unlikely to transfer to others and would give a false impression of generalisability. 
Learning rates, regularisation, label smoothing, and percentile clippings optimised for one configuration would likely not be optimal for another. Retuning them across configurations is computationally expensive, and they are unlikely to transfer well across surveys and instruments.
Table~\ref{tab:marginal} reports performance averaged over all configurations except one variable at a time. This provides a rough indication of which modelling choices, such as classifier architecture, image version, scale, or cropping strategy, that are most consistently good. This makes it a tentative guide for practitioners working on related datasets, even if the absolute performance values do not transfer directly. For instance, the DualSSN is for most purposes the best classifier. A more comprehensive version of Table~\ref{tab:marginal} would additionally average over hyperparameter choices such as learning rate, regularisation strength, and normalisation method, which were held fixed in this work. 

\section{Conclusions} \label{sec:conclusions}

This work investigated classification performance on the small but representative LoTSS-DR2/PSZ2 dataset as a step towards automated pipelines for detecting diffuse cluster radio emission in SKA surveys. Despite considerable source variance in angular extent, intensity, redshift, and morphology, several conclusions can be drawn.

\begin{enumerate}
    \item Dual-branch feature extraction consistently outperforms single-branch architectures on this small and complex dataset.
    \item Using the scattering coefficients in one of two branches, the DualSSN yields the best overall performance, and is the only architecture with stable accuracy across all cropping modes on reference data, i.e.\ the unprocessed LoTSS images without additional tapering, blurring, or point-source-subtraction.
    \item \textit{uv}-tapering, image-blurring, and point-source-subtraction all improve classifier performance relative to reference data, while stacking multiple image versions together as additional input channels does not. The comparable performance of blurred and \textit{uv}-tapered data indicates that image-domain blurring is a viable pre-processing alternative for visibility-free surveys.
    \item Beam-normalised cropping, which fixes the number of synthesised beams (and thus the noise correlation scale, i.e.\ the spatial extent over which neighbouring pixels are correlated) per image, outperforms fixed field-of-view and pixel cropping for blurred or \textit{uv}-tapered dataset versions. This suggests that a consistent noise scale is more important than a consistent angular size or minimal processing.
\end{enumerate}

Together, these results suggest that scattering-transform-based multi-branch architectures with beam-normalised cropping are a promising direction for diffuse emission classification in the SKA era.

Future work could test transfer learning for classification on X-ray images, and explore whether cluster mass, radius, or redshift provide complementary information. The noise stability was only evaluated on models trained on noisy data, but evaluating models trained on noiseless data on noisy data would be more informative on their noise stability. Extending the classification to distinguishing radio halos, relics, and phoenices is a natural next step as labelled samples grow. A vision transformer \citep{Sanvitale25} could serve as a third branch in a multi-branch classifier, and generative augmentation \citep{Mishra24} is an appealing strategy for expanding small training sets on LoTSS data.

\begin{acknowledgements}
MB acknowledges financial support from the SNSF under the Weave/Lead Agency project RadioClusters (214815). ET acknowledges financial support from the SNSF under the Starting Grant project Deep Waves (218396). The computations in this work were performed on the Alps supercomputer at the Swiss National Supercomputing Centre (CSCS), an HPE Cray EX system. We gratefully acknowledge the support provided by CSCS under project ID sk036, done in partnership with the SKACH consortium through funding by SERI. 
\end{acknowledgements}

\section*{Data availability}

Data from the LoTSS survey is available at \url{https://lofar-surveys.org/planck_dr2.html}. Code for models, processing, training, and plotting is available at \url{https://github.com/MarkusBredberg/dcreclass}.
 
\bibliographystyle{aa}
\bibliography{references}

\clearpage
\begin{appendix}
\nolinenumbers
\section{Architectures}\label{app:architectures}

Details about the architectures are given in this appendix. 
Tables~\ref{tab:CNN}--\ref{tab:DualScatterSqueezeNet} describe the 
architecture of the CNN, ScatterNet, DualCSN and DualSSN, respectively, 
using an image size of $128\times128$ and scattering coefficient size 
of $169\times32\times32$.

\begin{table}
    \centering
    \footnotesize
    \caption{CNN architecture.}
    \label{tab:CNN}
    \begin{tabular}{lcccc}
        \hline\hline
        Layer & Component & Depth & Act. & Reg. ($p$) \\
        \hline
        \multicolumn{5}{l}{\textit{Feature Extractor}} \\
        1--2   & $3\times3$ Conv, BN         &  8 & LReLU & DO2d (0.3) \\
        3--4   & $3\times3$ Conv, BN         & 16 & LReLU & DO2d (0.3) \\
        5      & $2\times2$ MP               & 16 & --    & --         \\
        6--7   & $5\times5$ Conv, BN         & 32 & LReLU & DO2d (0.4) \\
        8--9   & $3\times3$ Conv ($s$=2), BN & 32 & LReLU & DO2d (0.4) \\
        10--11 & $3\times3$ Conv ($s$=2), BN & 32 & LReLU & DO2d (0.4) \\
        12--13 & $3\times3$ Conv ($s$=2), BN & 32 & LReLU & DO2d (0.4) \\
        14--15 & $3\times3$ Conv ($s$=2), BN & 32 & LReLU & DO2d (0.4) \\
        \hline
        \multicolumn{5}{l}{\textit{Classifier}} \\
        1--2 & Linear, BN & 32 & LReLU & DO (0.5) \\
        3--4 & Linear, BN & 32 & LReLU & DO (0.5) \\
        5    & Linear     &  2 & --    & --       \\
        \hline
        \multicolumn{4}{r}{Total params:} & 69\,234 \\
        \hline
    \end{tabular}
    \tablefoot{LReLU: Leaky ReLU (slope 0.2). BN: Batch norm. MP: MaxPool. DO/DO2d: Dropout/Dropout2d. $s$: stride.}
\end{table}

\begin{table}
    \centering
    \footnotesize
    \caption{ScatterNet architecture.}
    \label{tab:ScatterNet}
    \begin{tabular}{lcccc}
        \hline\hline
        Layer & Component & Depth & Act. & Reg. ($p$) \\
        \hline
        \multicolumn{5}{l}{\textit{Classifier}} \\
        1 & Linear & 120 & ReLU & --       \\
        2 & Linear &  84 & ReLU & DO (0.5) \\
        3 & Linear &   2 & --   & --       \\
        \hline
        \multicolumn{4}{r}{Total params:} & 1\,976\,534 \\
        \hline
    \end{tabular}
    \tablefoot{DO: Dropout. See Table~\ref{tab:CNN} for other abbreviations.}
\end{table}

\begin{table}
    \centering
    \footnotesize
    \caption{DualConvolutionalSqueezeNet (DualCSN) architecture.}
    \label{tab:DualCNNSqueezeNet}
    \setlength{\tabcolsep}{3pt}
    \begin{tabular}{lcccc}
        \hline\hline
        Layer & Component & Depth & Act. & Reg. ($p$) \\
        \hline
        \multicolumn{5}{l}{\textit{Image Encoder 1}} \\
        1--2   & $5\times5$ Conv, BN                     &  8 & LReLU & DO2d (0.3)      \\
        3--4   & $3\times3$ Conv, BN                     & 16 & LReLU & DO2d (0.3)      \\
        5      & $2\times2$ MP                           & 16 & --    & --              \\
        6--7   & $5\times5$ Conv, BN                     & 32 & LReLU & DO2d (0.4)      \\
        8--9   & $3\times3$ Conv ($s$=2), BN             & 32 & LReLU & DO2d (0.4)      \\
        10--11 & $3\times3$ Conv ($s$=2), BN             & 32 & LReLU & DO2d (0.4)      \\
        12--15 & $2\times$ [$3\times3$ Conv ($s$=2), BN] & 32 & LReLU & DO2d (0.4, 0.5) \\
        \hline
        \multicolumn{5}{l}{\textit{Image Encoder 2}} \\
        1--2   & $3\times3$ Conv, BN, SE             & 32 & LReLU & DO2d (0.3) \\
        3--4   & $3\times3$ Conv, BN, SE             & 32 & LReLU & DO2d (0.3) \\
        5--6   & $3\times3$ Conv ($s$=2), BN, SE     & 32 & LReLU & DO2d (0.3) \\
        7--8   & $3\times3$ Conv, BN, SE             & 32 & LReLU & DO2d (0.4) \\
        9--10  & $3\times3$ Conv ($s$=2), BN, SE     & 32 & LReLU & DO2d (0.4) \\
        11--12 & $3\times3$ Conv, BN, SE             & 32 & LReLU & DO2d (0.5) \\
        13--14 & $3\times3$ Conv ($s$=2), BN, SE     & 32 & LReLU & DO2d (0.5) \\
        \hline
        \multicolumn{5}{l}{\textit{Classifier}} \\
        1--2 & Linear, BN & 32 & LReLU & DO (0.5) \\
        3--4 & Linear, BN & 32 & LReLU & DO (0.5) \\
        5    & Linear     &  2 & --    & --       \\
        \hline
        \multicolumn{4}{r}{Total params:} & 388\,530 \\
        \hline
    \end{tabular}
    \tablefoot{SE: Squeeze-and-excitation. See Table~\ref{tab:CNN} for other abbreviations.}
\end{table}

\begin{table}[ht!]
    \centering
    \footnotesize
    \caption{DualScatterSqueezeNet (DualSSN) architecture.}
    \label{tab:DualScatterSqueezeNet}
    \setlength{\tabcolsep}{3pt}
    \begin{tabular}{lcccc}
        \hline\hline
        Layer & Component & Depth & Act. & Reg. ($p$) \\
        \hline
        \multicolumn{5}{l}{\textit{Image Encoder}} \\
        1--2   & $5\times5$ Conv, BN                     &  8 & LReLU & DO2d (0.2)      \\
        3--4   & $3\times3$ Conv, BN                     & 16 & LReLU & DO2d (0.2)      \\
        5      & $2\times2$ MP                           & 16 & --    & --              \\
        6--7   & $5\times5$ Conv, BN                     & 32 & LReLU & DO2d (0.3)      \\
        8--9   & $3\times3$ Conv ($s$=2), BN             & 32 & LReLU & DO2d (0.3)      \\
        10--11 & $3\times3$ Conv ($s$=2), BN             & 32 & LReLU & DO2d (0.3)      \\
        12--15 & $2\times$ [$3\times3$ Conv ($s$=2), BN] & 32 & LReLU & DO2d (0.3, 0.4) \\
        \hline
        \multicolumn{5}{l}{\textit{Scattering Encoder}} \\
        1--2   & $3\times3$ Conv, BN, SE             & 16 & LReLU & DO2d (0.2) \\
        3--4   & $3\times3$ Conv, BN, SE             & 16 & LReLU & DO2d (0.2) \\
        5--6   & $3\times3$ Conv ($s$=2), BN, SE     & 16 & LReLU & DO2d (0.2) \\
        7--8   & $3\times3$ Conv, BN, SE             & 16 & LReLU & DO2d (0.3) \\
        9--10  & $3\times3$ Conv ($s$=2), BN, SE     & 16 & LReLU & DO2d (0.3) \\
        11--12 & $3\times3$ Conv, BN, SE             & 16 & LReLU & DO2d (0.4) \\
        13--14 & $3\times3$ Conv ($s$=2), BN, SE     & 16 & LReLU & DO2d (0.4) \\
        \hline
        \multicolumn{5}{l}{\textit{Classifier}} \\
        1--2 & Linear, BN & 32 & LReLU & DO (0.5) \\
        3--4 & Linear, BN & 32 & LReLU & DO (0.5) \\
        5    & Linear     &  2 & --    & --       \\
        \hline
        \multicolumn{4}{r}{Total params:} & 116\,146 \\
        \hline
    \end{tabular}
    \tablefoot{See Table~\ref{tab:CNN} and Table~\ref{tab:DualCNNSqueezeNet} for abbreviations.}
\end{table}

\FloatBarrier
\section{Image blurring and tapering} \label{app:blur}

$uv$-tapering is applied to the visibilities as part of the imaging process: taper weights are applied before the Fourier transform, so that the dirty image is already tapered when cleaning begins. Image-domain blurring, by contrast, is applied to the restored (post-deconvolution) reference image. Because $uv$-tapering modifies the synthesised beam and therefore the cleaning process, the two approaches are not equivalent, even when targeting the same physical scale.

If $\tilde I(u,v)$ denotes the Fourier transform of the sky brightness distribution $I(l,m)$, then
\[
I_{\mathrm{taper}}(l,m)
= \mathcal{F}^{-1}\{\tilde I(u,v)\,W(u,v)\} 
= I(l,m)\ast G_{\sigma_{\rm img}}(l,m),
\]
where $W(u,v)$ is a Gaussian \textit{uv}-taper
\[
W(u,v) = \exp\!\left[-\frac{u^{2}+v^{2}}{2\sigma_{uv}^{2}}\right],
\]
and $G_{\sigma_{\rm img}}(l,m) = \mathcal{F}^{-1}{W(u,v)}$ is the corresponding Gaussian convolution kernel in the image plane.
In this way, $uv$-space Gaussian tapering is mathematically equivalent to Gaussian convolution in the image plane, under the assumption of perfect $uv$-coverage. In practice, imperfect $uv$-coverage and the fact that blurring is applied after cleaning (rather than before, as in $uv$-tapering) mean the two diverge.

Because the reference image is already convolved with the synthesised beam (with covariance $\mathbf{C}_\mathrm{beam}$), an additional Gaussian convolution is required to reach the desired resolution. The covariance of this kernel is
\[
\mathbf{C}_\mathrm{ker} =
\mathbf{C}_\mathrm{target} - \mathbf{C}_\mathrm{beam},
\]
which is assumed to be positive definite so that the resulting beam is physically meaningful. The final tapered image is therefore
\[
I_T(\mathbf{x}) =
\left(I_\mathrm{ref} * \mathcal{G}_{\mathbf{C}_\mathrm{ker}}\right)(\mathbf{x}),
\]
where $\mathcal{G}_{\mathbf{C}_\mathrm{ker}}$ denotes a two-dimensional Gaussian with covariance matrix $\mathbf{C}_\mathrm{ker}$. The full procedure is shown in Algorithm~\ref{alg:blur}. The target beam is determined by the angular diameter distance because this method evaluates the visibility-free setting. There, redshift is the only available means of converting a desired physical resolution into an angular FWHM.

\begin{algorithm}
  \caption{\textsc{Blur}: convolve reference image to physical scale $s$.}
  \label{alg:blur}
  \begin{algorithmic}[1]
  \Require Reference image $I^{\rm Ref}$ with header $H^{\rm Ref}$;
           physical scale $s$ [kpc]; redshift $z$
  \Ensure Blurred image $I^{{\rm Blur}_s}$;
          effective beam FWHM $\theta^{\rm FWHM}_{{\rm Blur}_s}$
  \State $D_A\leftarrow\textsc{AngularDiameterDist}(z)$
  \State $\mathbf{C}_{\rm target}\leftarrow
         \left(\dfrac{s/D_A}{2\sqrt{2\ln 2}}\right)^{\!2}\mathbf{I}_2$
  \State $\mathbf{C}_{\rm beam}\leftarrow\textsc{BeamCovariance}(H^{\rm Ref})$
  \State $\mathbf{C}_{\rm ker}\leftarrow
         \mathrm{clip}_{\geq0}\!\bigl(\mathbf{C}_{\rm target}
         -\mathbf{C}_{\rm beam}\bigr)$
  \Comment{Remove negative eigenvalues}
  \State $K_s\leftarrow\textsc{Gaussian2D}(\mathbf{C}_{\rm ker})$
  \State $f_{\rm scale}\leftarrow
         2\pi\,\sigma_r^2 / \Omega^{\rm beam}(H^{\rm Ref})$
  \Comment{$\sigma_r = (s/D_A)/(2\sqrt{2\ln 2})$}
  \State $I^{{\rm Blur}_s}\leftarrow
         \bigl(I^{\rm Ref}\ast K_s\bigr)\cdot f_{\rm scale}$
  \State $\theta^{\rm FWHM}_{{\rm Blur}_s}\leftarrow
         \textsc{MajorAxisFWHM}\!\bigl(\mathbf{C}_{\rm beam}
         +\mathbf{C}_{\rm ker}\bigr)$
  \State\Return $I^{{\rm Blur}_s}$,\;$\theta^{\rm FWHM}_{{\rm Blur}_s}$
  \end{algorithmic}
\end{algorithm}

The angular diameter distance enters because, in a visibility-free setting, redshift is the only available means of converting a desired physical resolution (e.g. 50 kpc) into an angular FWHM for the reference image; no \textit{uv}-tapered counterpart is assumed to exist from which a target beam could otherwise be read.
  
Examples of \textit{uv}-tapered and blurred images are shown in Figure~\ref{fig:gaussian-blurring}. We see two major differences between the \textit{uv}-tapered images and our blurred images.

First, the blurred images are noisier than the \textit{uv}-tapered images. This is likely partly a consequence of the different sky regions in the versions combined with a per-image normalisation (see Section~\ref{sec:normalisation_and_augmentation}). If the \textit{uv}-tapered images contain a larger portion of high-intensity pixels, slightly brighter low-intensity pixels will be clipped to zero and the contrast will increase. Another possible explanation is that only \textit{uv}-tapering downweights the longer baseline and thus also the small-scale flux.

Second, the field of view varies slightly between the \textit{uv}-tapered and blurred images at the same nominal scale. Each image spans a fixed number of beam-widths, with the beam size specific to that image version. The effective beam size of the blurred image is the native-resolution beam convolved with the smoothing kernel, and is therefore generally different from the beam size of the \textit{uv}-tapered image at the same nominal physical scale. Because the two beam sizes differ, the same beam-normalised field of view corresponds to a different angular extent in each version. The number of beam-widths per image is set to be constant across all images with the same type and blurring scale so that the mean angular field of view is maximally equalised across all ten dataset versions. Any residual per-image variation is, thus, a consequence of the beam-size ratio varying with source redshift and native resolution. 

\begin{figure}
    \centering
    \includegraphics[width=\hsize]{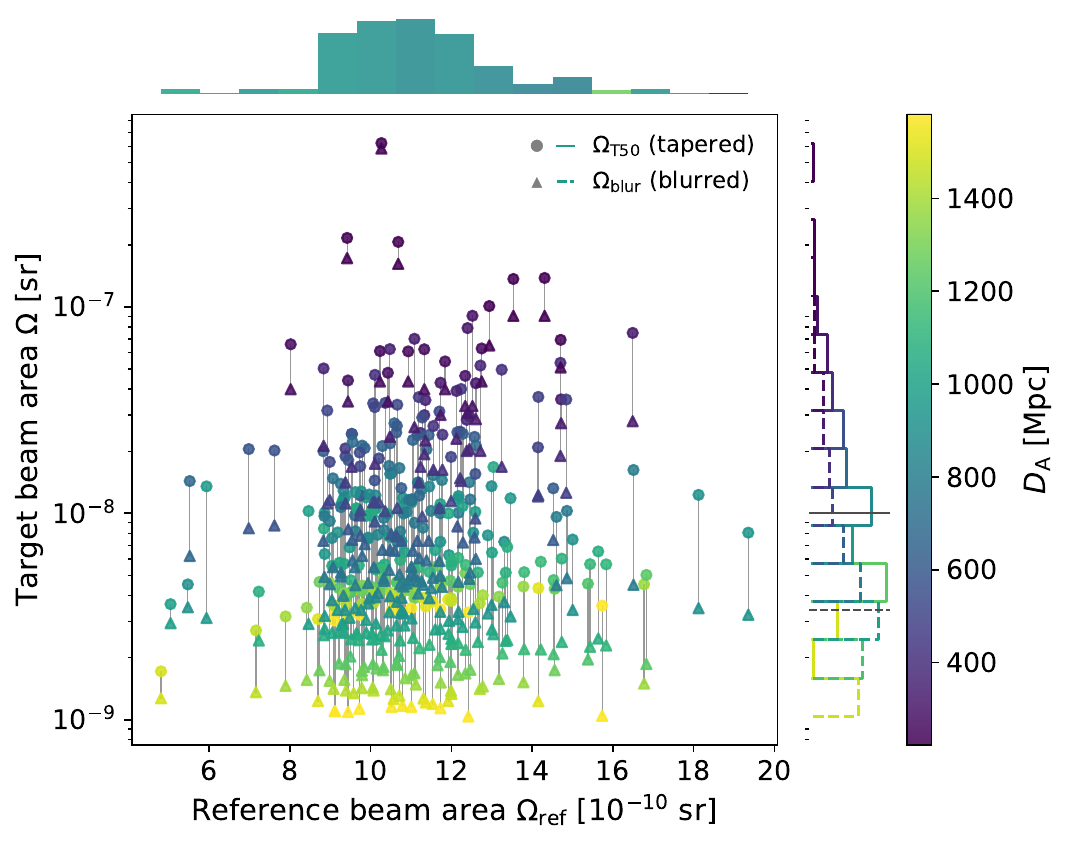}
    \caption{Scatter plot comparing the reference beam area against the 50$\,$kpc target beam area for each cluster image, for both \textit{uv}-tapered (circles) and image-blurred (triangles) versions. Colour indicates angular diameter distance. Vertical lines connect the two versions of the same cluster. Marginal histograms show the distribution of reference beam area (top) and target beam area (right), with solid and dashed lines marking the median for the tapered and blurred version, respectively.}
    \label{fig:scatter_taper}
\end{figure}

Figure~\ref{fig:scatter_taper} illustrates the difference in target beam size between the \textit{uv}-tapered and blurred images. The angular diameter distance decreases with an increasing tapering target beam area independently for the tapered and blurred images. Longer vertical lines connecting circles and triangles mean a larger difference in effective blurring scale. In all cases, the \textit{uv}-tapered image is blurred to a larger scale than the image-domain blurred counterpart. Another imperfection of the beam cropping is that not all effective beams are perfectly circular. Cropping the images according to a fixed number of beams across the width will therefore still allow for variation of number of beams in the height. Consequently, the noise will be a little bit less similar across images within a version than the method intends.

\section{Additional results} \label{sec:additional_results}

Table~\ref{tab:marginal} compares the mean performance per classifier, dataset version, blurring or tapering method, scale, and crop mode. Since the data and model are complex, the performance of each such variable is naturally dependent on the selection of the other parameters. This is partially reflected in the significant standard deviations, particularly for the recall. Still, some general conclusions can be drawn. For the classifiers, the DualSSN is generally performing better than the ScatterNet, which is systematically more stable than the CNN. The dataset version, scale, and crop mode are all resulting in, on average, small effects. Nevertheless, despite not creating the very best classifier in this study, point-source-subtracted dataset versions are often comparable or better than purely \textit{uv}-tapered or blurred data. A weak preference can also be found for \SI{25}{kpc} and beam cropping.

\begin{figure}
    \centering
    \includegraphics[width=\hsize]{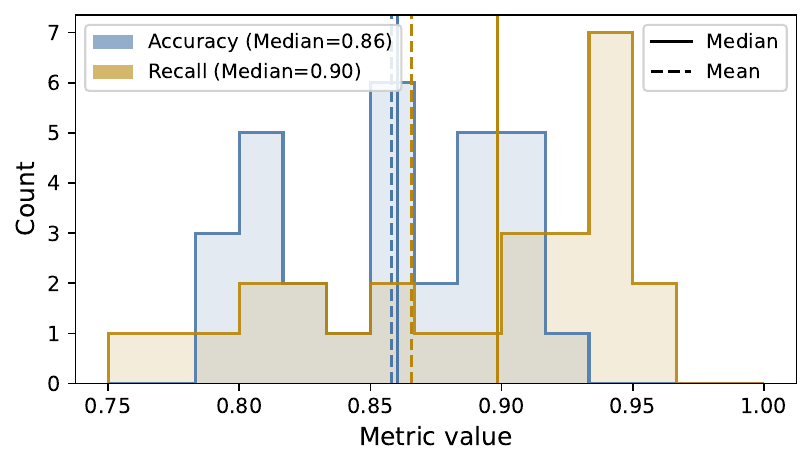}
    \caption{Histogram of accuracy (blue) and recall (gold) across the 30 runs for the best-performing configuration (DualSSN, beam-cropped \textit{uv}-tapered 25$\,$kpc data). Solid vertical lines mark the median; dashed lines mark the mean.}
    \label{fig:metric_dist}
\end{figure}

Figure~\ref{fig:metric_dist} shows the histogram for two selected metrics for the beam-cropped \SI{25}{kpc} \textit{uv}-tapered data classified with DualSSN. Since these metrics are roofed by 1.0, the median is often higher than the mean for well-performing sets of runs. Most importantly, the recall of this optimal configuration has a mean of 0.87 and a median of 0.90, with the bulk of all training runs resulting in a recall between 0.90 and 0.95.

This appendix also reports the full set of accuracies (Table~\ref{tab:acc_all_configs}), recalls (Table~\ref{tab:recall_all_configs}), F1-scores (Table~\ref{tab:f1_all_configs_transposed}), and AUC (Table~\ref{tab:auc_all_configs}) across all evaluated classifier architectures, dataset degradations, and cropping strategies. In these four metrics and ten different runs, the DualSSN with beam-cropped data is the best setup 28 out of 40 times. The second best, the DualCSN with beam cropping, only performs best four times out of 40.

\begin{table}
\centering
\small
\setlength{\tabcolsep}{3pt}
\caption{Performance statistics pooled across all complementary dimensions.}
\label{tab:marginal}
\begin{tabular}{lcccc}
\hline
Group & Accuracy & Recall & $F_1$ & AUC \\
\hline
\multicolumn{5}{l}{\textit{Best configuration: DualSSN, Tap 25\,kpc, Beam}} \\
\hline
 & $\mathbf{0.86 \pm 0.04}$ & $\mathbf{0.87 \pm 0.08}$ & $\mathbf{0.85 \pm 0.05}$ & $\mathbf{0.93 \pm 0.04}$ \\
\hline
\multicolumn{5}{l}{\textit{Per classifier}} \\
\hline
CNN & $0.67 \pm 0.10$ & $0.72 \pm 0.25$ & $0.64 \pm 0.20$ & $0.74 \pm 0.13$ \\
ScatterNet & $0.62 \pm 0.07$ & $0.63 \pm 0.18$ & $0.59 \pm 0.11$ & $0.66 \pm 0.08$ \\
DualCSN & $0.68 \pm 0.09$ & $\mathbf{0.77 \pm 0.15}$ & $0.68 \pm 0.10$ & $0.75 \pm 0.09$ \\
DualSSN & $\mathbf{0.74 \pm 0.10}$ & $0.76 \pm 0.17$ & $\mathbf{0.72 \pm 0.12}$ & $\mathbf{0.82 \pm 0.10}$ \\
\hline
\multicolumn{5}{l}{\textit{Per version}} \\
\hline
Reference & $0.66 \pm 0.08$ & $0.71 \pm 0.18$ & $0.65 \pm 0.13$ & $0.73 \pm 0.10$ \\
25\,kpc Tap & $0.71 \pm 0.12$ & $0.77 \pm 0.18$ & $\mathbf{0.70 \pm 0.15}$ & $0.77 \pm 0.14$ \\
25\,kpc Blur & $0.67 \pm 0.11$ & $0.68 \pm 0.21$ & $0.63 \pm 0.18$ & $0.71 \pm 0.13$ \\
25\,kpcTap+Sub & $\mathbf{0.71 \pm 0.08}$ & $0.76 \pm 0.17$ & $0.70 \pm 0.12$ & $\mathbf{0.80 \pm 0.09}$ \\
50\,kpc Tap & $0.67 \pm 0.10$ & $0.70 \pm 0.20$ & $0.65 \pm 0.15$ & $0.73 \pm 0.13$ \\
50\,kpc Blur & $0.71 \pm 0.11$ & $0.72 \pm 0.19$ & $0.68 \pm 0.16$ & $0.76 \pm 0.12$ \\
50\,kpcTap+Sub & $0.66 \pm 0.09$ & $\mathbf{0.82 \pm 0.19}$ & $0.68 \pm 0.13$ & $0.75 \pm 0.10$ \\
100\,kpc Tap & $0.66 \pm 0.10$ & $0.62 \pm 0.19$ & $0.61 \pm 0.15$ & $0.72 \pm 0.13$ \\
100\,kpc Blur & $0.63 \pm 0.07$ & $0.63 \pm 0.19$ & $0.59 \pm 0.14$ & $0.68 \pm 0.08$ \\
100\,kpcTap+Sub & $0.68 \pm 0.10$ & $0.82 \pm 0.16$ & $0.69 \pm 0.11$ & $0.79 \pm 0.08$ \\
\hline
\multicolumn{5}{l}{\textit{Per type}} \\
\hline
Tap & $0.68 \pm 0.11$ & $0.70 \pm 0.20$ & $0.65 \pm 0.16$ & $0.74 \pm 0.13$ \\
Blur & $0.67 \pm 0.10$ & $0.67 \pm 0.20$ & $0.64 \pm 0.16$ & $0.72 \pm 0.12$ \\
Tap + Sub & $\mathbf{0.68 \pm 0.09}$ & $\mathbf{0.80 \pm 0.18}$ & $\mathbf{0.69 \pm 0.12}$ & $\mathbf{0.78 \pm 0.09}$ \\
\hline
\multicolumn{5}{l}{\textit{Per scale}} \\
\hline
25\,kpc & $\mathbf{0.69 \pm 0.11}$ & $0.74 \pm 0.19$ & $\mathbf{0.68 \pm 0.16}$ & $\mathbf{0.76 \pm 0.13}$ \\
50\,kpc & $0.68 \pm 0.10$ & $\mathbf{0.75 \pm 0.20}$ & $0.67 \pm 0.15$ & $0.75 \pm 0.12$ \\
100\,kpc & $0.66 \pm 0.09$ & $0.69 \pm 0.20$ & $0.63 \pm 0.14$ & $0.73 \pm 0.11$ \\
\hline
\multicolumn{5}{l}{\textit{Per crop mode}} \\
\hline
Beam crop & $\mathbf{0.69 \pm 0.11}$ & $\mathbf{0.74 \pm 0.19}$ & $\mathbf{0.68 \pm 0.15}$ & $\mathbf{0.77 \pm 0.12}$ \\
FoV crop & $0.66 \pm 0.09$ & $0.71 \pm 0.21$ & $0.64 \pm 0.14$ & $0.72 \pm 0.10$ \\
\hline
\end{tabular}
\tablefoot{Each row reports mean $\pm$ standard deviation, fixing one variable (classifier, image version, physical scale, or crop mode) and averaging over all other configurations. Pixel cropping results are not included. Tap: Gaussian \textit{uv}-taper applied in the visibility plane. Blur: Gaussian convolution applied in the image plane to approximate \textit{uv}-tapering. Sub: Point-source-subtracted image at the corresponding \textit{uv}-tapered scale. Beam: beam-normalised cropping. FoV: fixed field-of-view cropping. Best result within each section in \textbf{bold}.}
\end{table}

\definecolor{cnncol}{RGB}{245,248,255}
\definecolor{scatcol}{RGB}{248,255,245}
\definecolor{dualcsncol}{RGB}{255,248,245}
\definecolor{dualssncol}{RGB}{250,245,255}
\definecolor{rowgray}{gray}{0.96}
\arrayrulecolor{gray!40}
\begin{table*}
\centering
\caption{Accuracy for all classifier--version--crop configurations.}
\label{tab:acc_all_configs}
\rowcolors{2}{rowgray}{white}
\resizebox{\textwidth}{!}{%
\begin{tabular}{l c c c c c c c c c c}
\hline
 & Reference
 & 25\,kpc Blur
 & 25\,kpc Tap.
 & 25\,kpc Tap. + Sub.
 & 50\,kpc Blur
 & 50\,kpc Tap.
 & 50\,kpc Tap. + Sub.
 & 100\,kpc Blur
 & 100\,kpc Tap.
 & 100\,kpc Tap. + Sub. \\
\hline
CNN Beam
& $0.67 \pm 0.06$ & $0.71 \pm 0.08$ & $0.72 \pm 0.13$ & $0.72 \pm 0.04$
& $0.68 \pm 0.10$ & $0.77 \pm 0.11$ & $0.65 \pm 0.10$
& $0.64 \pm 0.06$ & $0.78 \pm 0.03$ & $0.63 \pm 0.13$ \\
CNN FoV
& $0.62 \pm 0.07$ & $0.61 \pm 0.08$ & $0.66 \pm 0.10$ & $0.69 \pm 0.07$
& $0.73 \pm 0.10$ & $0.65 \pm 0.08$ & $0.68 \pm 0.08$
& $0.58 \pm 0.04$ & $0.60 \pm 0.04$ & $0.67 \pm 0.09$ \\
CNN Pix.
& $0.65 \pm 0.14$ & $0.58 \pm 0.06$ & $0.62 \pm 0.09$ & $0.66 \pm 0.08$
& --- & --- & ---
& --- & --- & --- \\
ScatterNet Beam
& $0.62 \pm 0.03$ & $0.55 \pm 0.06$ & $0.58 \pm 0.06$ & $0.68 \pm 0.06$
& $0.56 \pm 0.04$ & $0.55 \pm 0.04$ & $0.62 \pm 0.03$
& $\mathbf{0.67 \pm 0.04}$ & $0.57 \pm 0.03$ & $0.71 \pm 0.05$ \\
ScatterNet FoV
& $0.60 \pm 0.04$ & $0.60 \pm 0.07$ & $0.61 \pm 0.05$ & $0.67 \pm 0.04$
& $0.62 \pm 0.08$ & $0.60 \pm 0.05$ & $0.65 \pm 0.03$
& $0.64 \pm 0.04$ & $0.54 \pm 0.03$ & $0.69 \pm 0.02$ \\
ScatterNet Pix.
& $0.67 \pm 0.09$ & $0.59 \pm 0.04$ & $0.57 \pm 0.04$ & $0.66 \pm 0.04$
& --- & --- & ---
& --- & --- & --- \\
DualCSN Beam
& $0.64 \pm 0.08$ & $0.70 \pm 0.06$ & $0.80 \pm 0.07$ & $0.71 \pm 0.08$
& $0.74 \pm 0.06$ & $0.71 \pm 0.10$ & $0.62 \pm 0.08$
& $0.64 \pm 0.08$ & $\mathbf{0.78 \pm 0.04}$ & $0.65 \pm 0.09$ \\
DualCSN FoV
& $0.66 \pm 0.07$ & $0.66 \pm 0.07$ & $0.69 \pm 0.09$ & $0.66 \pm 0.07$
& $0.73 \pm 0.09$ & $0.68 \pm 0.05$ & $0.64 \pm 0.07$
& $0.59 \pm 0.06$ & $0.64 \pm 0.03$ & $0.65 \pm 0.08$ \\
DualCSN Pix.
& $0.77 \pm 0.08$ & $0.59 \pm 0.09$ & $0.62 \pm 0.07$ & $0.60 \pm 0.09$
& --- & --- & ---
& --- & --- & --- \\
DualSSN Beam
& $0.76 \pm 0.06$ & $\mathbf{0.79 \pm 0.08}$ & $\mathbf{0.86 \pm 0.04}$ & $\mathbf{0.81 \pm 0.10}$
& $\mathbf{0.81 \pm 0.04}$ & $\mathbf{0.78 \pm 0.07}$ & $\mathbf{0.74 \pm 0.10}$
& $0.67 \pm 0.06$ & $0.77 \pm 0.03$ & $\mathbf{0.77 \pm 0.07}$ \\
DualSSN FoV
& $0.74 \pm 0.06$ & $0.73 \pm 0.09$ & $0.73 \pm 0.08$ & $0.75 \pm 0.06$
& $0.80 \pm 0.04$ & $0.67 \pm 0.06$ & $0.68 \pm 0.10$
& $0.61 \pm 0.08$ & $0.60 \pm 0.07$ & $0.64 \pm 0.11$ \\
DualSSN Pix.
& $\mathbf{0.82 \pm 0.03}$ & $0.68 \pm 0.08$ & $0.68 \pm 0.11$ & $0.74 \pm 0.10$
& --- & --- & ---
& --- & --- & --- \\
\hline
\end{tabular}}
\tablefoot{Best result per column in \textbf{bold}.}
\end{table*}

\begin{table*}
\centering
\caption{Recall for all classifier--version--crop configurations.}
\label{tab:recall_all_configs}
\rowcolors{2}{rowgray}{white}
\resizebox{\textwidth}{!}{%
\begin{tabular}{l c c c c c c c c c c}
\hline
 & Reference
 & 25\,kpc Blur
 & 25\,kpc Tap.
 & 25\,kpc Tap. + Sub.
 & 50\,kpc Blur
 & 50\,kpc Tap.
 & 50\,kpc Tap. + Sub.
 & 100\,kpc Blur
 & 100\,kpc Tap.
 & 100\,kpc Tap. + Sub. \\
\hline
CNN Beam
& $0.75 \pm 0.16$ & $0.65 \pm 0.26$ & $0.78 \pm 0.26$ & $0.79 \pm 0.15$
& $0.73 \pm 0.27$ & $\mathbf{0.84 \pm 0.23}$ & $0.85 \pm 0.29$
& $0.60 \pm 0.23$ & $0.73 \pm 0.12$ & $\mathbf{0.89 \pm 0.18}$ \\
CNN FoV
& $0.68 \pm 0.21$ & $0.64 \pm 0.25$ & $0.78 \pm 0.25$ & $0.79 \pm 0.22$
& $0.71 \pm 0.22$ & $0.68 \pm 0.27$ & $0.82 \pm 0.28$
& $0.52 \pm 0.21$ & $0.48 \pm 0.19$ & $0.78 \pm 0.27$ \\
CNN Pix.
& $0.70 \pm 0.27$ & $0.62 \pm 0.29$ & $0.74 \pm 0.24$ & $0.77 \pm 0.20$
& --- & --- & ---
& --- & --- & --- \\
ScatterNet Beam
& $0.61 \pm 0.22$ & $0.55 \pm 0.20$ & $0.64 \pm 0.17$ & $0.62 \pm 0.18$
& $0.58 \pm 0.18$ & $0.59 \pm 0.17$ & $0.68 \pm 0.17$
& $0.67 \pm 0.10$ & $0.56 \pm 0.13$ & $0.73 \pm 0.09$ \\
ScatterNet FoV
& $0.59 \pm 0.19$ & $0.61 \pm 0.21$ & $0.67 \pm 0.19$ & $0.64 \pm 0.13$
& $0.61 \pm 0.24$ & $0.59 \pm 0.17$ & $0.68 \pm 0.12$
& $\mathbf{0.72 \pm 0.15}$ & $0.65 \pm 0.19$ & $0.72 \pm 0.05$ \\
ScatterNet Pix.
& $0.72 \pm 0.11$ & $0.63 \pm 0.15$ & $0.61 \pm 0.15$ & $0.56 \pm 0.14$
& --- & --- & ---
& --- & --- & --- \\
DualCSN Beam
& $\mathbf{0.79 \pm 0.16}$ & $0.72 \pm 0.15$ & $0.84 \pm 0.05$ & $0.79 \pm 0.16$
& $0.78 \pm 0.08$ & $0.83 \pm 0.19$ & $0.88 \pm 0.11$
& $0.63 \pm 0.25$ & $\mathbf{0.75 \pm 0.10}$ & $0.84 \pm 0.09$ \\
DualCSN FoV
& $0.76 \pm 0.10$ & $0.74 \pm 0.16$ & $0.81 \pm 0.11$ & $0.82 \pm 0.08$
& $0.79 \pm 0.09$ & $0.70 \pm 0.09$ & $0.88 \pm 0.09$
& $0.66 \pm 0.17$ & $0.57 \pm 0.11$ & $0.88 \pm 0.08$ \\
DualCSN Pix.
& $0.75 \pm 0.08$ & $0.75 \pm 0.20$ & $0.75 \pm 0.12$ & $0.75 \pm 0.13$
& --- & --- & ---
& --- & --- & --- \\
DualSSN Beam
& $0.78 \pm 0.13$ & $\mathbf{0.81 \pm 0.06}$ & $\mathbf{0.87 \pm 0.08}$ & $\mathbf{0.87 \pm 0.12}$
& $\mathbf{0.81 \pm 0.09}$ & $0.77 \pm 0.10$ & $0.87 \pm 0.11$
& $0.66 \pm 0.12$ & $0.71 \pm 0.17$ & $0.85 \pm 0.10$ \\
DualSSN FoV
& $0.75 \pm 0.13$ & $0.70 \pm 0.21$ & $0.78 \pm 0.11$ & $0.81 \pm 0.11$
& $0.77 \pm 0.08$ & $0.64 \pm 0.13$ & $\mathbf{0.91 \pm 0.08}$
& $0.56 \pm 0.20$ & $0.49 \pm 0.24$ & $0.86 \pm 0.19$ \\
DualSSN Pix.
& $0.74 \pm 0.06$ & $0.68 \pm 0.15$ & $0.77 \pm 0.13$ & $0.82 \pm 0.15$
& --- & --- & ---
& --- & --- & --- \\
\hline
\end{tabular}}
\tablefoot{Best result per column in \textbf{bold}.}
\end{table*}

\begin{table*}
\centering
\caption{$F_1$ scores for all classifier--version--crop configurations.}
\label{tab:f1_all_configs_transposed}
\rowcolors{2}{rowgray}{white}
\resizebox{\textwidth}{!}{%
\begin{tabular}{l c c c c c c c c c c}
\hline
 & Reference
 & 25\,kpc Blur
 & 25\,kpc Tap.
 & 25\,kpc Tap. + Sub.
 & 50\,kpc Blur
 & 50\,kpc Tap.
 & 50\,kpc Tap. + Sub.
 & 100\,kpc Blur
 & 100\,kpc Tap.
 & 100\,kpc Tap. + Sub. \\
\hline
CNN Beam
& $0.66 \pm 0.13$
& $0.63 \pm 0.24$
& $0.70 \pm 0.22$
& $0.71 \pm 0.13$
& $0.64 \pm 0.23$
& $0.74 \pm 0.21$
& $0.65 \pm 0.21$
& $0.57 \pm 0.20$
& $0.75 \pm 0.06$
& $0.68 \pm 0.14$ \\
CNN FoV
& $0.60 \pm 0.16$
& $0.56 \pm 0.20$
& $0.65 \pm 0.19$
& $0.67 \pm 0.18$
& $0.69 \pm 0.19$
& $0.60 \pm 0.21$
& $0.66 \pm 0.22$
& $0.51 \pm 0.15$
& $0.50 \pm 0.17$
& $0.65 \pm 0.22$ \\
CNN Pix.
& $0.54 \pm 0.19$
& $0.52 \pm 0.23$
& $0.62 \pm 0.17$
& $0.66 \pm 0.14$
& ---
& ---
& ---
& ---
& ---
& --- \\
ScatterNet Beam
& $0.58 \pm 0.12$
& $0.51 \pm 0.13$
& $0.57 \pm 0.11$
& $0.63 \pm 0.11$
& $0.53 \pm 0.12$
& $0.54 \pm 0.10$
& $0.61 \pm 0.09$
& $\mathbf{0.65 \pm 0.05}$
& $0.54 \pm 0.06$
& $0.70 \pm 0.03$ \\
ScatterNet FoV
& $0.56 \pm 0.11$
& $0.56 \pm 0.16$
& $0.60 \pm 0.12$
& $0.63 \pm 0.08$
& $0.57 \pm 0.17$
& $0.56 \pm 0.10$
& $0.64 \pm 0.06$
& $0.64 \pm 0.05$
& $0.56 \pm 0.07$
& $0.68 \pm 0.03$ \\
ScatterNet Pix.
& $0.59 \pm 0.04$
& $0.57 \pm 0.09$
& $0.56 \pm 0.09$
& $0.59 \pm 0.08$
& ---
& ---
& ---
& ---
& ---
& --- \\
DualCSN Beam
& $0.66 \pm 0.13$
& $0.68 \pm 0.13$
& $0.80 \pm 0.06$
& $0.70 \pm 0.14$
& $0.73 \pm 0.04$
& $0.71 \pm 0.15$
& $0.68 \pm 0.05$
& $0.58 \pm 0.20$
& $\mathbf{0.76 \pm 0.05}$
& $0.69 \pm 0.04$ \\
DualCSN FoV
& $0.67 \pm 0.03$
& $0.65 \pm 0.12$
& $0.71 \pm 0.05$
& $0.69 \pm 0.03$
& $0.73 \pm 0.05$
& $0.67 \pm 0.03$
& $0.69 \pm 0.03$
& $0.59 \pm 0.05$
& $0.59 \pm 0.06$
& $0.70 \pm 0.03$ \\
DualCSN Pix.
& $0.68 \pm 0.06$
& $0.62 \pm 0.13$
& $0.64 \pm 0.06$
& $0.63 \pm 0.04$
& ---
& ---
& ---
& ---
& ---
& --- \\
DualSSN Beam
& $\mathbf{0.75 \pm 0.05}$
& $\mathbf{0.78 \pm 0.05}$
& $\mathbf{0.85 \pm 0.05}$
& $\mathbf{0.81 \pm 0.08}$
& $\mathbf{0.80 \pm 0.04}$
& $\mathbf{0.76 \pm 0.06}$
& $\mathbf{0.76 \pm 0.08}$
& $0.65 \pm 0.04$
& $0.73 \pm 0.08$
& $\mathbf{0.77 \pm 0.04}$ \\
DualSSN FoV
& $0.72 \pm 0.07$
& $0.68 \pm 0.19$
& $0.73 \pm 0.07$
& $0.74 \pm 0.05$
& $0.78 \pm 0.04$
& $0.64 \pm 0.09$
& $0.73 \pm 0.05$
& $0.55 \pm 0.14$
& $0.50 \pm 0.18$
& $0.68 \pm 0.13$ \\
DualSSN Pix.
& $0.72 \pm 0.04$
& $0.66 \pm 0.10$
& $0.69 \pm 0.06$
& $0.74 \pm 0.11$
& ---
& ---
& ---
& ---
& ---
& --- \\
\hline
\end{tabular}}
\tablefoot{Best result per column in \textbf{bold}.}
\end{table*}

\begin{table*}
\centering
\caption{AUC for all classifier--version--crop configurations.}
\label{tab:auc_all_configs}
\rowcolors{2}{rowgray}{white}
\resizebox{\textwidth}{!}{%
\begin{tabular}{l c c c c c c c c c c}
\hline
 & Reference
 & 25\,kpc Blur
 & 25\,kpc Tap.
 & 25\,kpc Tap. + Sub.
 & 50\,kpc Blur
 & 50\,kpc Tap.
 & 50\,kpc Tap. + Sub.
 & 100\,kpc Blur
 & 100\,kpc Tap.
 & 100\,kpc Tap. + Sub. \\
\hline
CNN Beam & $0.74 \pm 0.07$ & $0.76 \pm 0.11$ & $0.80 \pm 0.16$ & $0.83 \pm 0.08$ & $0.73 \pm 0.14$ & $0.84 \pm 0.13$ & $0.75 \pm 0.13$ & $0.67 \pm 0.08$ & $\mathbf{0.88 \pm 0.04}$ & $0.77 \pm 0.14$ \\
CNN FoV & $0.69 \pm 0.10$ & $0.67 \pm 0.11$ & $0.73 \pm 0.16$ & $0.77 \pm 0.10$ & $0.78 \pm 0.10$ & $0.68 \pm 0.11$ & $0.76 \pm 0.13$ & $0.61 \pm 0.06$ & $0.63 \pm 0.07$ & $0.79 \pm 0.07$ \\
CNN Pix. & $0.74 \pm 0.12$ & $0.64 \pm 0.10$ & $0.71 \pm 0.12$ & $0.76 \pm 0.12$ & --- & --- & --- & --- & --- & --- \\
ScatterNet Beam & $0.67 \pm 0.03$ & $0.56 \pm 0.04$ & $0.64 \pm 0.07$ & $0.75 \pm 0.06$ & $0.61 \pm 0.04$ & $0.59 \pm 0.04$ & $0.68 \pm 0.02$ & $0.73 \pm 0.04$ & $0.61 \pm 0.02$ & $0.80 \pm 0.07$ \\
ScatterNet FoV & $0.63 \pm 0.04$ & $0.60 \pm 0.05$ & $0.63 \pm 0.04$ & $0.71 \pm 0.03$ & $0.65 \pm 0.05$ & $0.62 \pm 0.03$ & $0.71 \pm 0.02$ & $0.68 \pm 0.05$ & $0.57 \pm 0.04$ & $0.75 \pm 0.03$ \\
ScatterNet Pix. & $0.73 \pm 0.03$ & $0.62 \pm 0.04$ & $0.57 \pm 0.03$ & $0.70 \pm 0.02$ & --- & --- & --- & --- & --- & --- \\
DualCSN Beam & $0.74 \pm 0.09$ & $0.75 \pm 0.08$ & $0.89 \pm 0.03$ & $0.81 \pm 0.07$ & $0.81 \pm 0.07$ & $0.80 \pm 0.12$ & $0.74 \pm 0.08$ & $0.69 \pm 0.09$ & $0.86 \pm 0.03$ & $0.77 \pm 0.08$ \\
DualCSN FoV & $0.72 \pm 0.07$ & $0.72 \pm 0.10$ & $0.76 \pm 0.10$ & $0.74 \pm 0.04$ & $0.78 \pm 0.09$ & $0.72 \pm 0.05$ & $0.74 \pm 0.06$ & $0.62 \pm 0.06$ & $0.67 \pm 0.03$ & $0.76 \pm 0.06$ \\
DualCSN Pix. & $0.84 \pm 0.05$ & $0.66 \pm 0.10$ & $0.69 \pm 0.09$ & $0.73 \pm 0.05$ & --- & --- & --- & --- & --- & --- \\
DualSSN Beam & $0.85 \pm 0.05$ & $\mathbf{0.88 \pm 0.05}$ & $\mathbf{0.93 \pm 0.04}$ & $\mathbf{0.93 \pm 0.05}$ & $\mathbf{0.89 \pm 0.04}$ & $\mathbf{0.86 \pm 0.08}$ & $\mathbf{0.85 \pm 0.09}$ & $\mathbf{0.74 \pm 0.07}$ & $0.88 \pm 0.02$ & $\mathbf{0.87 \pm 0.03}$ \\
DualSSN FoV & $0.82 \pm 0.06$ & $0.78 \pm 0.12$ & $0.82 \pm 0.09$ & $0.84 \pm 0.06$ & $0.85 \pm 0.03$ & $0.73 \pm 0.06$ & $0.79 \pm 0.08$ & $0.68 \pm 0.08$ & $0.64 \pm 0.09$ & $0.78 \pm 0.07$ \\
DualSSN Pix. & $\mathbf{0.87 \pm 0.03}$ & $0.75 \pm 0.07$ & $0.74 \pm 0.14$ & $0.84 \pm 0.10$ & --- & --- & --- & --- & --- & --- \\
\hline
\end{tabular}}
\tablefoot{Best result per column in \textbf{bold}.}
\end{table*}

\end{appendix}

\end{document}